\documentclass[3p]{elsarticle}
\biboptions{sort&compress}
\usepackage{graphicx}
\usepackage{epsfig}
\usepackage{times}
\usepackage{color}
\usepackage{amsmath,amssymb}
\begin{document}
\begin{frontmatter}
\title{Non-Lagrangian approach for coupled complex Ginzburg-Landau systems with higher order-dispersion}
\author[uoy1]{Roger Bertin Djob\corref{cor1}}
\ead{rogerdjob@yahoo.fr}
\author[uoy2]{Aurelien Kenfact-Jiotsa}
\author[cnld]{A. Govindarajan}
\ead{govin.nld@gmail.com}
\cortext[cor1]{Corresponding author}
\address[uoy1]{Laboratory of Mechanics, Group of Nonlinear
    Physics and Complex Systems, Department of Physics,
                  Faculty of Sciences, University of Yaounde I, P.O. Box 812, Yaounde, Cameroon}
 \address[uoy2]{Nonlinear Physics and Complex Systems Group,
                Department of Physics, Higher Teacher's Training College,
                       University of Yaounde I, P.O. Box 47 Yaounde, Cameroon}
 \address[cnld]{Centre for Nonlinear Dynamics, School of Physics, Bharathidasan University, Tiruchirappalli - 620 024, India}
\begin{abstract}
It is known that after a particular distance of evolution in fiber
lasers, two (input) asymmetric soliton like pulses emerge as two
(output) symmetric pulses having same and constant energy. We
report such a compensation technique in dispersion managed fiber
lasers by means of a semi-analytical method known as collective
variable  approach (CVA) with including third-order dispersion
(TOD). The minimum length of fiber laser, at which the output
symmetric pulses are obtained from the input asymmetric ones, is
calculated for each and every pulse parameters numerically by
employing  Runge-Kutta fourth order method. The impacts of  intercore linear
coupling,  asymmetric nature of initial parameters and TOD on the
evolution of  pulse parameters and on the minimum length are also
investigated. It is found that strong intercore linear coupling and asymmetric nature of input pulse parameters result
in the reduction of fiber laser length. Also, the role of TOD tends to increase the
width of the pulses as well as their energies. Besides, chaotic
patterns and bifurcation points on the minimum length of the fiber
owing to the impact of TOD are also reported in a nutshell.
\end{abstract}
\end{frontmatter}
\section{INTRODUCTION}
In nonlinear science, complex Ginzburg-Landau equations (CGLEs)
serve as well-known models to represent ubiquitous nonlinear
dynamics representing pattern formation and solitons for a number
of media, which include reaction-diffusion systems, hydrodynamics,
plasma systems, nonlinear optics and so forth
\cite{aranson2002world,boris2005,winful1992passive,
arecchi1999pattern,ipsen2000amplitude,fermann1997fiber,he2007fusion,liu2011phase,lai2018generation}.
In the context of nonlinear optics, the CGLE is obtained as a
perturbation to the standard nonlinear Schrodinger equation (NLSE)
for a system of fiber laser (of single and coupled versions)
taking into account gain, loss and spectral filtering
\cite{moubissi2001non}. For a detailed review, readers can refer
to \cite{malomed2007solitary,malomed2017variety}. Among different
types of fiber lasers,  dual-core ring lasers have been receiving
a widespread interest owing to their numerous applications in
photonics
\cite{malomed1996stable,atai1996stability,sakaguchi1995phase,sakaguchi1996traveling,sakaguchi2001instabilities,atai1998exact,sakaguchi2000breathing}.
Nonlinear directional couplers (NLDC) are one of important
components of integrated optical devices and they deliver quite a
few lightwave applications including switching (bright and dark)
\cite{jensen1982nonlinear,govind2014femtosecond,govind2015numerical,govind2014dark,sarma2009dark,sarma2011vector,govindarajan2019nonlinear,govindarajan2019tailoring},
logical operations \cite{govindaraji2016interaction,correia2017obtaining}, ultra-short pulse
generation through modulational instability
\cite{ganapathy2006modulational,govindarajan2017modulational,li2011mod,li2012mod,kominis2017stability,zhiyenbayev2019enhanced}, and
amplifiers \cite{malomed1996nonlinear,chu1997pulse}. In
particular, Winful and Walton have proven that the asymmetric dual
core fibers can act as a passive mode-locking fiber laser provided
one of the dual-cores must be doped with rare earth-elements like
erbium while the other core must be a passive one
\cite{winful1992passive}. Such a device has successfully
demonstrated a soliton generation with a huge amplification
\cite{walton1993passive}.  Following the former work, Kanka has
numerically proposed a more realistic fiber laser employing
Lorentzian resonant gain profile including stimulated Raman
scattering \cite{kavnka1994numerical}. A ring type fiber laser has
been put forward by Oh \textit{et al.} by considering higher order
dispersion such as third-order dispersion and it has been reported
that TOD and self-steepening do not alter the soliton generation
substantially \cite{oh1995robust}.

Along similar lines, quite a few theoretical studies have been
emerged in the literature in order to identify the stability
nature of such solitons. Especially, Malomed and Winful have
reported the first ever stable solitons in the coupled fiber
systems though the earlier observed one was completely unstable
\cite{malomed1996stable}. In addition, stable bound non-stationary
solitons were also identified in the framework of coupled CGL
systems \cite{sigler2005solitary}. Moreover, the instability
conditions and nonlinear development of modulational instability
were also reported
\cite{ganapathy2006modulational,porsezian2009modulational,alcaraz2010modulational}.
These studies employing nonlinear waves, especially solitary
waves, clearly exhibit that such fiber lasers based on coupled
CGLEs can manifest a number of applications in lightwave
communication systems \cite{malomed2007solitary}. Meanwhile, the
study of dispersion management in optical fibers receives
considerable interest. It is important to note that instead of
using fibers with only anomalous dispersion as in case of soliton
mode locking, dispersion managed solitons are generated by
utilizing a dispersion map where  normal and anomalous dispersions
in fibers are periodically displaced
\cite{berntson1999dispersion}.

The utilization of dispersion map is usually referred as
dispersion management (DM) and the pertaining solutions are
further referred as DM solitons. In this technique, the dispersion
parameter is related to the GVD by the following relation; since
the dispersion is a function of propagation distance, the pulse
experiences periodic net balance between dispersion and
nonlinearity at different positions and thus the pulse duration
and chirp oscillate periodically along the propagation, which is
why DM solitons are also called \textit{stretched pulses}.  On average, the
pulse in the dispersion map broadens due to the net anomalous
dispersion which should be exactly balanced by the nonlinearity
and thus it goes back to its initial pulse width and spectrum
bandwidth after each round-trip propagation
\cite{malomed2006soliton}. Thus these breathing pulses can still
be viewed as solitons on the average sense and are good candidate
for mode locked fiber lasers since the periodic cavity boundary
condition can be easily satisfied. Interestingly, these stretched
pulses can also be generated from a cavity with small net normal
dispersion. DM solitons have drawn significant attentions since
the alternating sign of dispersions (normal and anomalous) helps
suppressing the nonlinear phase accumulation which results in
higher energies, usually by one order of magnitude, in comparison
to the conventional soliton lasers. Besides, the breathing
property of DM solitons helps to suppress the Gordon- Haus timing
jitter in optical communication systems \cite{wong2001dispersion}.

It is important to consider the higher order effects such as
third-order dispersion (TOD), self-steepening and stimulated Raman
scattering if the width of the pulse is considered to be 100 fs or
even below
\cite{fewo2005dispersion,fewo2008collective,biswas2008femtosecond}.
In particular, the TOD parameter is more important in optical
communications and one should consider when the pump wavelength is
very close to zero group velocity dispersion.  The importance of
TOD term in models of fiber lasers has been well described in the
literature. To enumerate them, Fewo \textit{et al}. have
investigated the role of TOD parameter in fiber amplifiers by
means of collective variable analysis with the inclusion of
quintic nonlinearity.  They have also illustrated the influence of
the third order dispersion and the cubic-quintic (CQ) nonlinear
coefficients on the evolution of soliton pulses for different
pulse parameters. Quite recently, Sakaguchi \textit{et al.} have
shown that the model of fiber-laser cavities near the zero
dispersion point, based on the Complex Ginzburg-Landau equation
with the cubic-quintic nonlinearity and TOD term, supports stable
dissipative solitons and demonstrated that the same model gives
rise to several specific families of robust bound states of
solitons \cite{sakaguchi2018stationary}.

Recently, we have highlighted the interaction between two Gaussian
pulses propagating in such a fiber laser by the means of
collective variables approach \cite{djob2015study}. This
interaction was governed by a pair of CQ-CGLE without higher order
dispersion (HOD) terms. The main result was the compensation of
the energies between both cores whatever the input amplitudes and
widths. Along these directions, one of the basic and necessary
problems is to generate soliton sources with suitable physical and
practical characteristics, such as compact and stable source of
interest, compact size, and commercially available one
\cite{green2008dynamics}. Nevertheless, no study is focused on the
dispersion management  technique of dual-core fiber ring lasers
based on the TOD effect and to determine the \textit{minimum
distance} after which the energies of input asymmetric pulses in
neighbouring cores become constant and equal each other. Indeed,
the issue of symmetric output pulses is relevant as the equal
energy in the neighbouring cores of the fiber avoid compensation,
no fluctuations and the energy of the emerging pulses can easily
be evaluated. Motivated by all these facts, in this paper, we
present a comprehensive picture of TOD term on the pulse shaping
of optical solitons and identify the minimum length at which one
obtains the symmetric bright solitons pair in the proposed system.
To fulfil this goal, the paper is organized as follows. In Sec. 2,
we derive the equations of motion from the CVs approach and
briefly explain the dispersion management technique leading to the
propagation of conventional solitons. In Sec. 3, we present a
semi-analytical approach, which leads to determine the minimum
distance or minimum length of the fiber for obtaining symmetric
output pulses. Finally, in Sec. 4 we conclude 
the paper.
\section{Theoretical model and analytical techniques}
\label{sec:2}
The theoretical model of the present work is based on the linearly
coupled CQ-CGLEs with the third order dispersion parameter, which can be written as
\begin{equation}\label{eq1}
\begin{split}
&i\Psi_{1,x}+(p_r+ip_i)\Psi_{1,tt}+(q_{r}+iq_{i})|\Psi_1|^2\Psi_1=i(\gamma_r+i \gamma_i)\Psi_1+(c_r+ic_i)|\Psi_1|^4\Psi_1+i(d_r+id_i)\Psi_{1,ttt}+k\Psi_2\\
&i\Psi_{2,x}+(p_r+ip_i)\Psi_{2,tt}+(q_{r}+iq_{i})|\Psi_2|^2\Psi_2=i(\gamma_r+i\gamma_i)\Psi_2+(c_r+ic_i)|\Psi_2|^4\Psi_2+i(d_r+id_i)\Psi_{2,ttt}+k\Psi_1,
\end{split}
\end{equation}
where $\Psi_{i}$ ($i=1,2$) are normalized (in soliton units)
complex envelopes of the field components (ie., $\Psi_1$ is the
propagating field inside the core 1 and $\Psi_2$ refers to the
field inside the core 2) and $t$ and $x$ are, respectively, the
normalized distance and the retarded time. Also, the parameters
$p_r$, $p_i$, $d_r$, $d_i$, $q_{r}$, $q_{i}$, $c_{r}$, $c_{i}$,
$\gamma_r$ and $ \gamma_i$ are real constants and in what follows,
without loss of generality, they can be commonly expressed as a
function of $x$ without loosing their constant characters, viz.,
$p_r = p_r(x)$, $p_i=p_i(x)$, $d_r = d_r(x)$, $d_i=d_i(x)$, $q_r =
q_r(x)$, $q_i= q_i(x)$, $c_r = c_r(x)$, $c_i=c_i(x)$, $ \gamma_r =
\gamma_r(x)$, and $\gamma_i= \gamma_i(x)$, respectively. Further,
it is to be noted that the term $p_r$ measures the wave dispersion
while $p_i$ denotes the spectral filtering coefficient. Likewise,
$d_r$ is the TOD coefficient, $d_i$ the cubic frequency dependent
gain/loss coefficient, $q_{r}$ and $q_{i}$ accordingly represent
the nonlinear coefficient and nonlinear gain-absorption
parameters. It is worthwhile to mention that the nonlinear gain
helps to suppress the growth of radiative background (linear mode)
which always accompanies the propagation of nonlinear stationary
pulses in real fiber links. Furthermore, $c_{r}$ and $c_{i}$ stand
for the higher-order correction terms to the nonlinear refractive
index and nonlinear amplification absorption, respectively. The
other terms $ \gamma_r$ and $ \gamma_i$ represent the linear gain
and frequency shift, respectively whereas the inter core linear
coupling coefficient is noted through the parameter $k$.

\subsection{Collective variables approach}\label{2}
In the context of nonlinear fiber optics, there exist a few analytical methods to study the dynamics of optical pulses. If we consider the effects of optical losses, and higher order dispersion including non-paraxial limit, the non-integrable nature tend to make them hard to study analytically and one has to seek the aid of relevant numerical methods \cite{tamil}. Nevertheless, a non-Lagrangian method, namely collective variable  approach (CVA) can be employed to any perturbative NLSE as it does not require any conserved energy to reduce the governing equations \cite{veljkovic2015super,veljkovic2016super,asma2019chirped}. Hence, it is  possible to derive the CV equations for the perturbative NLSE which makes them very useful to study the former. It is well-known that the collective variable approach  is
mainly dependent on the separation of original applied fields $\Psi_j$, ($j=1, 2$) into
some appropriate functions known as ansatzes ($f_1$ and $f_2$)
with residual fields ($g_1$ and $g_2$) as
\begin{equation}\label{eq2}
\begin{split}
&\Psi_1(x,t)=f_1(u_1,....,u_N,t)+g_1(x,t)\\
&\Psi_2(x,t)=f_2(v_1,....,v_N,t)+g_2(x,t),
\end{split}
\end{equation}
where $u_i$ and $v_i$ are space-dependent variables and can
describe the nature of applied fields. Let us now consider the
Gaussian ansatzes, which read as
\begin{equation}\label{eq15}
\begin{split}
& f_1(u_1,u_2,{u_3},{u_4},{u_5},{u_6},t)=u_1\,{{\rm e}^{-{\frac {
\left( t-u_2 \right) ^{2}}{u_3^{2}}}}}{{\rm e}^{
\,i\frac{{u_4}}{2}\, \left( t-u_2 \right) ^{2}+i{ u_5}\,
\left(t-u_2
\right) +i{u_6}}}\\
&f_2(v_1,v_2,{v_3},{v_4},{v_5},v_6,t)=v_1\,{{\rm e}^{-{\frac {
\left( t-v_2 \right) ^{2}}{v_3^{2}}}}}{{\rm e}^{
\,i\frac{{v_4}}{2}\, \left( t-v_2 \right) ^{2}+i{ v_5}\,
\left(t-v_2 \right) +iv_6}}.
\end{split}
\end{equation}
In Eq. \eqref{eq15}, the collective variables $u_1$, $u_2$,
$\sqrt{2\ln(2)}{u_3}$, $\frac{{u_4}}{2\pi}$, $\frac{u_5}{2\pi}$
and ${u_6}$ accordingly indicate the pulse amplitude, temporal
position, pulse width (FWHM), chirp, frequency and phase of the
pulse propagating in the first core, while other terms such as
$v_1$, $v_2$, $\sqrt{2\ln(2)}{v_3}$, $\frac{{v_4}}{2\pi}$,
$\frac{{v_5}}{2\pi}$ and $v_6$ refer the same as defined in the
above for the second core of the fiber laser.

To determine the role each collective variables, one has to reduce the number of terms in the residual field energies. The expressions of
the residual field energies are written
\begin{equation}\label{eq3}
\begin{split}
&\varepsilon_1=\int_ {-\infty}^ {+\infty} |g_1(x,t)|^{2}\,dt =
\int_
{-\infty}^ {+\infty} |\Psi_1(x,t)-f_1(u_i,t)|^{2}\,dt\\
&\varepsilon_2=\int_ {-\infty}^ {+\infty} |g_2(x,t)|^{2}\,dt =
\int_ {-\infty}^ {+\infty}
|\Psi_2(x,t)-f_2(v_i,t)|^{2}\,dt,i=1..N;
\end{split}
\end{equation}
 Also, the total energy calculated in each of cores can be further read as
\begin{equation}\label{eq4}
\begin{split}
&E_1=\int_ {-\infty}^ {+\infty} |\Psi_1(x,t)|^{2}\,dt\\
&E_2=\int_ {-\infty}^ {+\infty} |\Psi_2(x,t)|^{2}\,dt.
\end{split}
\end{equation}
Hence, the expressions for the calculated energy of  the residual fields become
\begin{equation}\label{eq5}
\begin{split}
&\varepsilon_1=\int_ {-\infty}^ {+\infty} |f_1(u_i,t)|^{2}\,dt  -
2R_e(\int_ {-\infty}^ {+\infty} \Psi_1(x,t) f_1^{*}(u_i,t)\,dt)\\
&\varepsilon_2=\int_ {-\infty}^ {+\infty} |f_2(v_i,t)|^{2}\,dt  -
2R_e(\int_ {-\infty}^ {+\infty} \Psi_2(x,t) f_2^{*}(v_i,t)\,dt),
\end{split}
\end{equation}
Note that $^{*}$ indicates the complex conjugate.

If one knows  the ansatzes that must satisfy the minimization of
the residual field energies, we can then obtain the first
constraint equations of the form
\begin{equation}\label{eq6}
\begin{split}
&C_{i}^1=\frac{\partial \varepsilon_1}{\partial u_i}\\
&C_{i}^2=\frac{\partial \varepsilon_2}{\partial v_i}.
\end{split}
\end{equation}
which can also be read as
\begin{equation}\label{eq7}
\begin{split}
&C_{i}^1=2R_e( - \int_ {-\infty}^ {+\infty} g_1f^{*}_{1,u_i}\,dt)\\
&C_{i}^2=2R_e( - \int_ {-\infty}^ {+\infty}g_2f^{*}_{2,v_i}\,dt),
\end{split}
\end{equation}
where $f^{}_{1,u_i}$ and $f^{}_{2,v_i}$ indicate, respectively, the first derivatives of
$f_1$ and $f_2$ with respect to pulse parameters $u_i$ and $v_i$. Similarly, the second constraint equations of motions read as
\begin{equation}\label{eq8}
\begin{split}
&\dot{C_{i}^1}=0\\
 &\dot{C_{i}^2}=0;
 \end{split}
\end{equation}
where the terms $\dot{C_{i}^1}$ and $\dot{C_{i}^2}$ are, respectively, the first derivatives with
respect to $x$ of $C_{i}^1$ and $C_{i}^2$. Proceeding further, the second constraint equations are written as
\begin{equation}\label{eq9}
\begin{split}
&\dot{C_{i}^1}=2R_e[ - \int_ {-\infty}^ {+\infty} g_1
f^{*}_{1,u_iu_j}\,dt- \int_ {-\infty}^ {+\infty} g_{1,x}
f^{*}_{1,u_i}\,dt]=0\\
&\dot{C_{i}^2}=2R_e[ - \int_ {-\infty}^ {+\infty} g_2
f^{*}_{2,v_iv_j}\,dt- \int_ {-\infty}^ {+\infty} g_{2,x}
f^{*}_{2,v_i}\,dt]=0;
\end{split}
\end{equation}
Here, the term $f_{x_i}$ is the first derivative with respect to $x_i$ of $f$.
Employing the \textit{bare approximation} i.e., $g\approx0$ and considering the slow
variations of the residual fields with the time viz.,  $g_{tt}\approx0$,
the constraint equations become further
\begin{equation} \label{eq10}
\begin{split}
 \dot{C_{i}^1}=&2R_e[\sum(<f^{*}_{1,u_i}\cdot f_{1,u_j}>-<f^{*}_{1,u_i}\cdot i(p_r+ip_i)\cdot f_{1,tt}>+<f^{*}_{1,u_i}\cdot (d_r+id_i)\cdot f_{1,ttt}>  \\
   &-<f^{*}_{1,u_i}\cdot (\ \gamma_r+i\ \gamma_i)\cdot f_1>-<f^{*}_{1,u_i}\cdot
   i(q_{r}+iq_{i})\cdot|f_1|^2f_1> \\
&+<f^{*}_{1,u_i}\cdot i(c_{r}+ic_{i})\cdot
|f_1|^4f_1>+<f^{*}_{1,v_i}\cdot k \cdot f_2>]=0\\
\dot{C_{i}^2}=&2R_e[\sum(<f^{*}_{2,v_i}\cdot f_{2,v_j}>-<f^{*}_{2,v_i}\cdot i(p_r+ip_i)\cdot f_{2,tt}>+<f^{*}_{2,u_i}\cdot (d_r+id_i)\cdot f_{2,ttt}>\\
   &-<f^{*}_{2,v_i}\cdot (\ \gamma_r+i\ \gamma_i)\cdot f_2>-<f^{*}_{2,v_i}\cdot
   i(q_{r}+iq_{i})\cdot|f_2|^2f_2> \\
&+<f^{*}_{v_i}\cdot i(c_{r}+ic_{i})\cdot
|f_2|^4f_2>+<f^{*}_{2,v_i}\cdot k \cdot f_1>]=0,
\end{split}
 \end{equation}
In all the above equations, the terms indicated in the closed
parenthesis indicate the integration with respect to the time as,
$ <(...)>= \int_ {-\infty}^ {+\infty} (...)\,dt $ and $R_e(...)$
refers to the real part of the complex $(...)$. The final
constraint equations of motion for the pulse parameters can be
read as
\begin{equation}\label{eq11}
\begin{split}
&[\dot{C^1}]=[\frac{\partial C^1}{\partial U}][\dot{U}]+[R^1]=0\\
&[\dot{C^2}]=[\frac{\partial C^2}{\partial V}][\dot{V}]+[R^2]=0,
\end{split}
\end{equation}
which lead to,
\begin{equation}\label{eq12}
\begin{split}
&[\dot{U}]=-[\frac{\partial C^1}{\partial U}]^{-1}[R^1]\\
&[\dot{V}]=-[\frac{\partial C^2}{\partial V}]^{-1}[R^2],
\end{split}
\end{equation}
where $[\frac{\partial C^1}{\partial U}]$ and $[\frac{\partial
C^2}{\partial V}]$ are square matrices whose elements are noted
\begin{equation}\label{eq13}
\begin{split}
&C_{ij}^1=-2R_e(<f^{*}_{1,u_i}\cdot f_{u_j}>)\\
&C_{ij}^2=-2R_e(<f^{*}_{2,v_i}\cdot f_{v_j}>).
\end{split}
\end{equation}
While $[\frac{\partial C^1}{\partial U}]^{-1}$ and
$[\frac{\partial C^2}{\partial V}]^{-1}$ are inverse matrices of
$[\frac{\partial C^1}{\partial U}]$ and $[\frac{\partial
C^2}{\partial V}]$, $[R^{1,2}_i]$ are column vectors whose elements  are
written as
\begin{equation} \label{eq14}
\begin{split}
 R_{i}^1=&2R_e(-<f^{*}_{1,u_i}\cdot i(p_r+ip_i)\cdot f_{1,tt}>+<f^{*}_{1,u_i}\cdot (d_r+id_i)\cdot f_{1,ttt}>-<f^{*}_{1,u_i}\cdot (\ \gamma_r+i\ \gamma_i)\cdot f_1>\\
   &-<f^{*}_{1,u_i}\cdot i(q_{r}+iq_{i})\cdot |f_1|^2f_1>+<f^{*}_{1,u_i}\cdot i(c_{r}+ic_{i})\cdot|f_1|^4f_1>)+<f^{*}_{1,u_i}\cdot k \cdot f_2>)\\
   R_{i}^2=&2R_e(-<f^{*}_{2,v_i}\cdot i(p_r+ip_i)\cdot f_{2,tt}>+<f^{*}_{2,u_i}\cdot (d_r+id_i)\cdot f_{2,ttt}>-<f^{*}_{2,v_i}\cdot (\ \gamma_r+i\ \gamma_i)\cdot f_2>\\
   &-<f^{*}_{2,v_i}\cdot i(q_{r}+iq_{i})\cdot |f_2|^2f_2>+<f^{*}_{2,v_i}\cdot i(c_{r}+ic_{i})\cdot|f_2|^4f_2>)+<f^{*}_{2,v_i}\cdot k \cdot f_1>).
\end{split}
 \end{equation}

Applying all the above developments, one obtains the equations of
motion of the pulse parameters as given in Appendix A
 \subsection{The dispersion management technique}
\begin{figure}[ht]
\centering
    \includegraphics[width=0.3\linewidth]{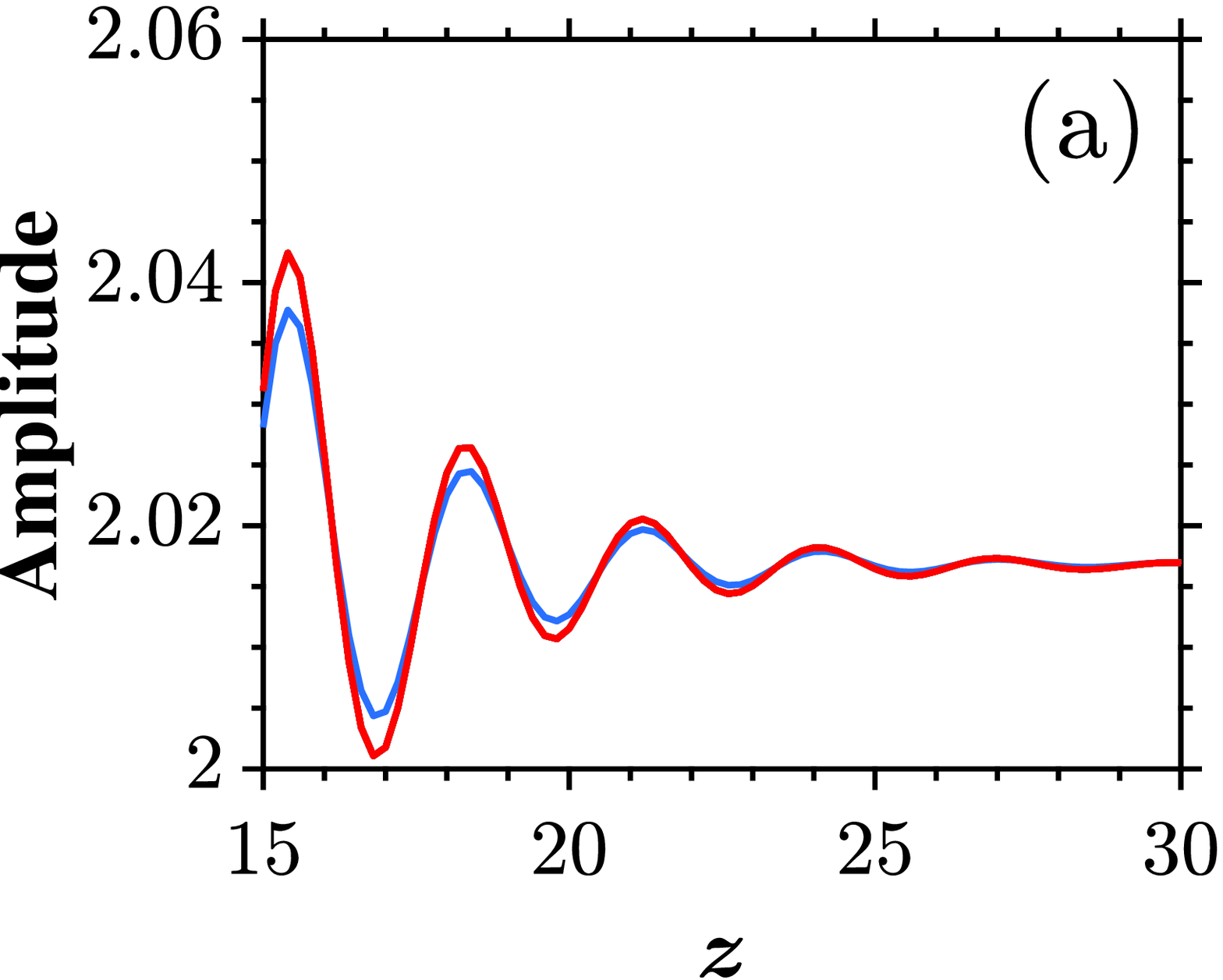}\includegraphics[width=0.3\linewidth]{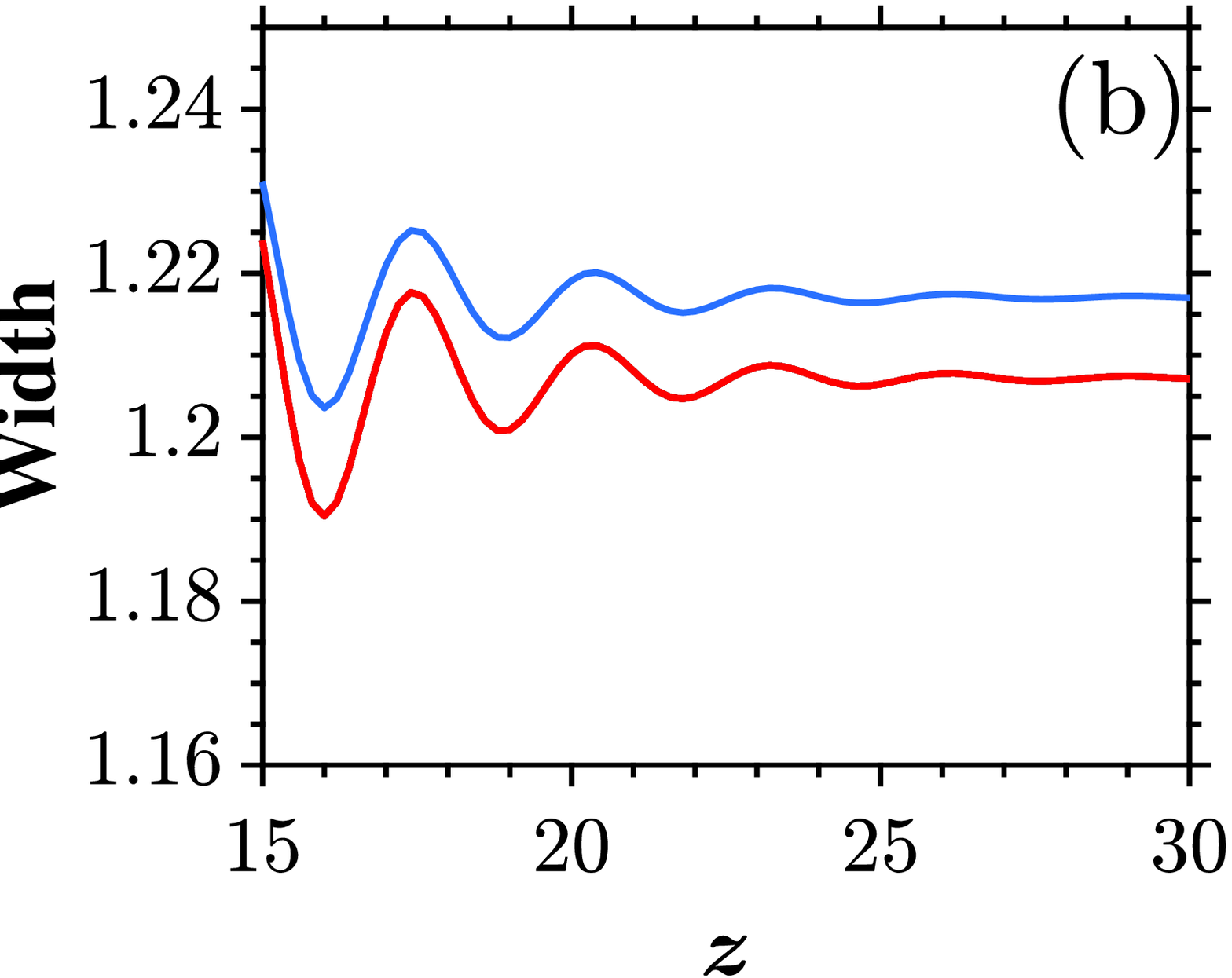}\\
     \includegraphics[width=0.62\linewidth]{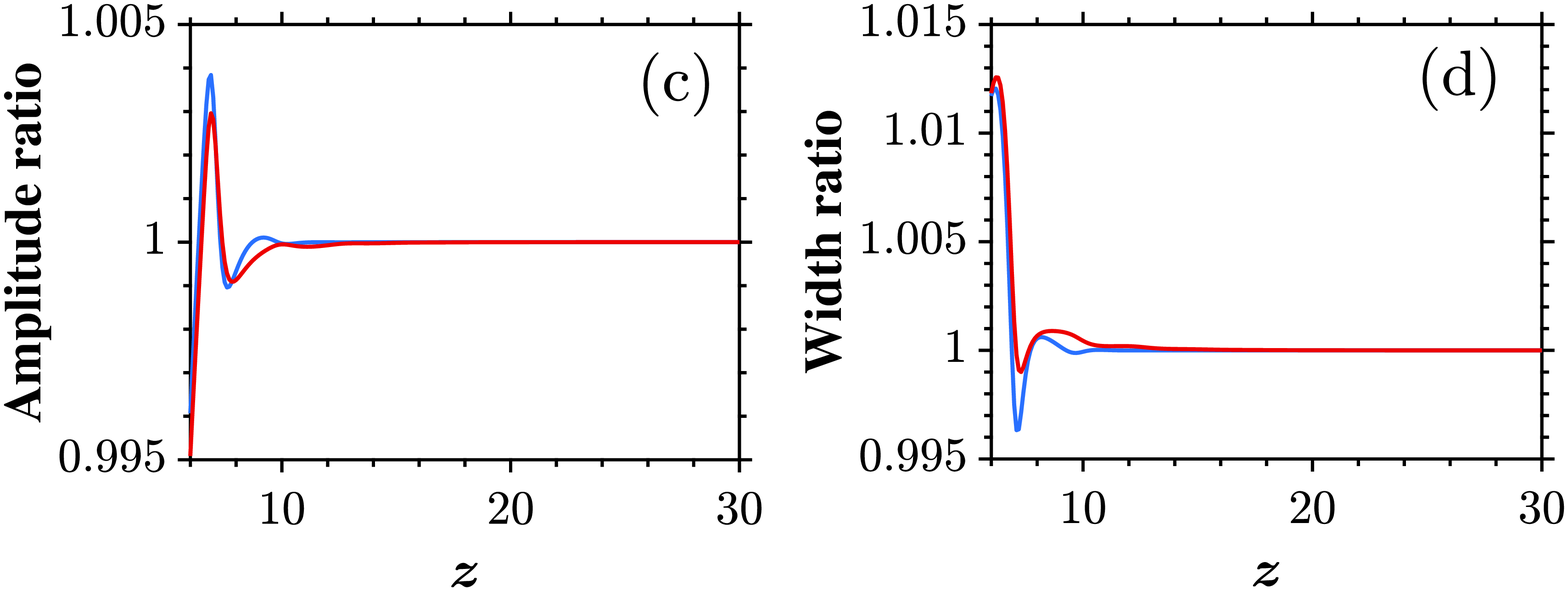}
\caption{\emph{ {\small Effect of TOD on the spatial evolution of
amplitudes and widths of input asymmetric gaussian pulses in the
first core of fiber laser (the dynamics in the second core is not shown as
it exhibits almost the same dynamics as in the first core ). The
blue curves correspond to $d_{r}=0.0$ and the red curves are drawn
with $d_{r}=-0.1$. The initial parameters values are chosen such
that $u_{10}=u_{30}=1$, $v_{10}=0.75$, $v_{30}=1$ and the coupling
coefficient value is to be,  $k=0.5$.}}}\label{fig22}
\end{figure}
 The dispersion management technique aims to form non-conventional soliton by
alternating normal ($p_r<0$) and anomalous ($p_r>0$) dispersions
and assuring overall zero dispersion. Its principle consists of
anomalous-dispersion ($p_r=d_1>0$) section with a length $z_1$,
followed by a normal-dispersion ($p_r=d_2<0$) section with a
length $z_2$ respecting the zero-average dispersion as written in the following form
\begin{equation} \label{c2eq27b}
d_1z_1+d_2z_2\approx 0
 \end{equation}
 It must be noted that in this work, we fix the length of fiber laser as $z_1=0.025$, $z_2=0.025$
(corresponding to the anomalous and normal dispersion regimes)
and the group velocity dispersion coefficient to be $p_r=0.5$ and
$p_r=-0.5$ for the anomalous and normal dispersion regimes,
respectively. The numerical study pertaining to the evolution of
neighboring pulse parameters along the propagation distance is
carried out using the standard  Runge--Kutta fourth order method
with a spatial step size of $2\times10^{-3}$.
\section{ Results and discussions} \label{3}
\subsection{Spatial evolution of input asymmetric pulses}
\begin{figure}[ht]
\centering
\includegraphics[width=0.6\linewidth]{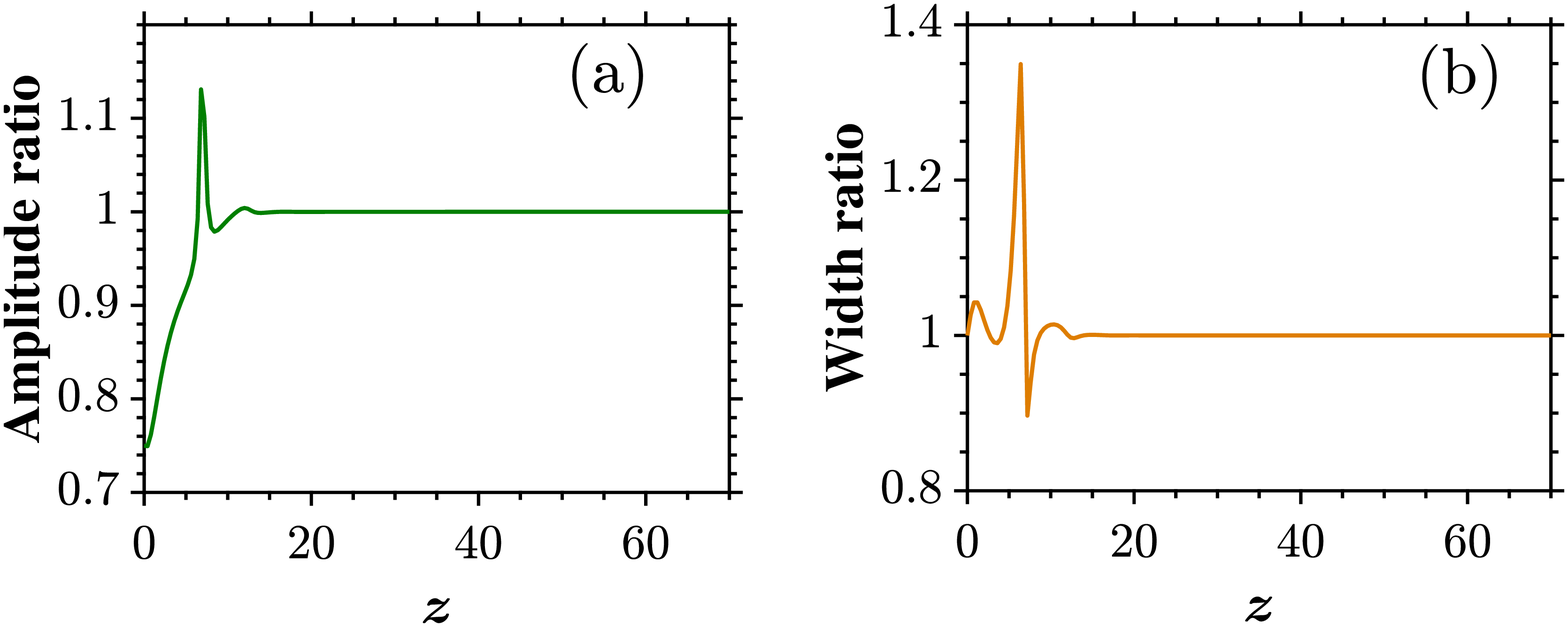}\\
    \includegraphics[width=0.6\linewidth]{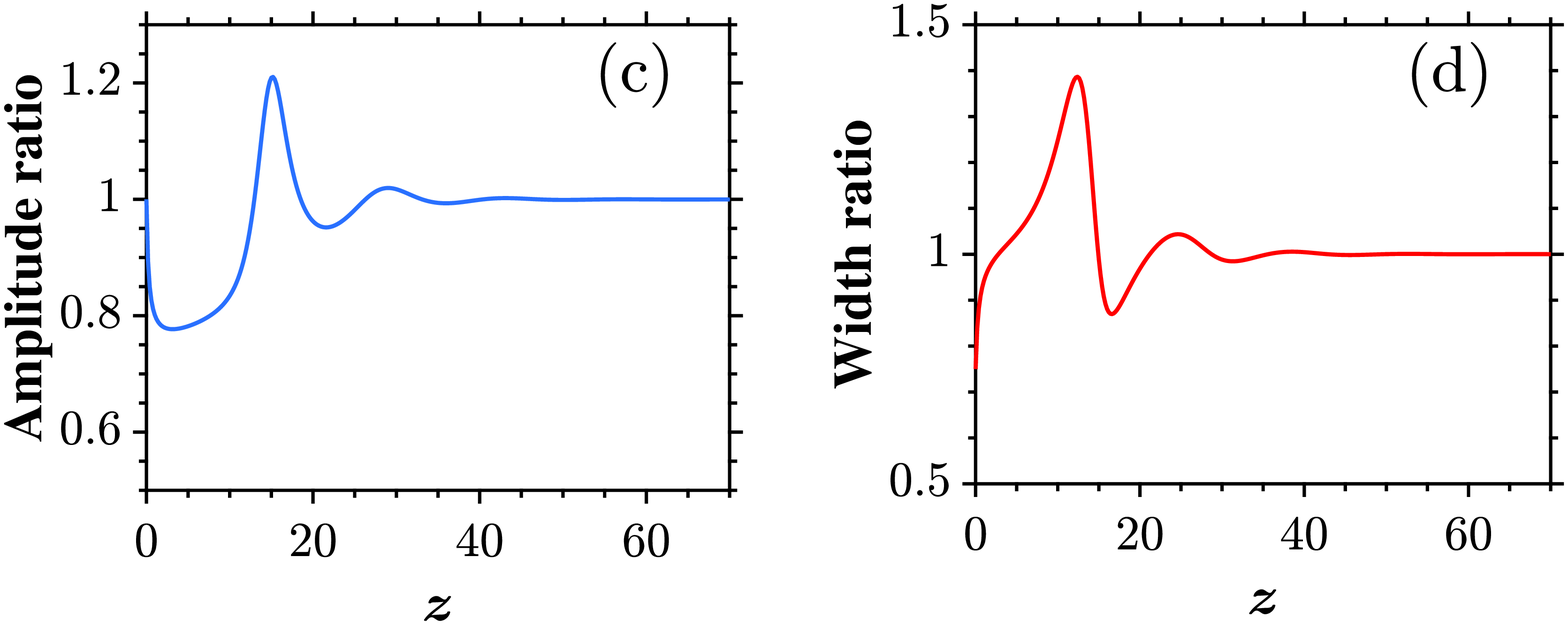}\\
    \includegraphics[width=0.6\linewidth]{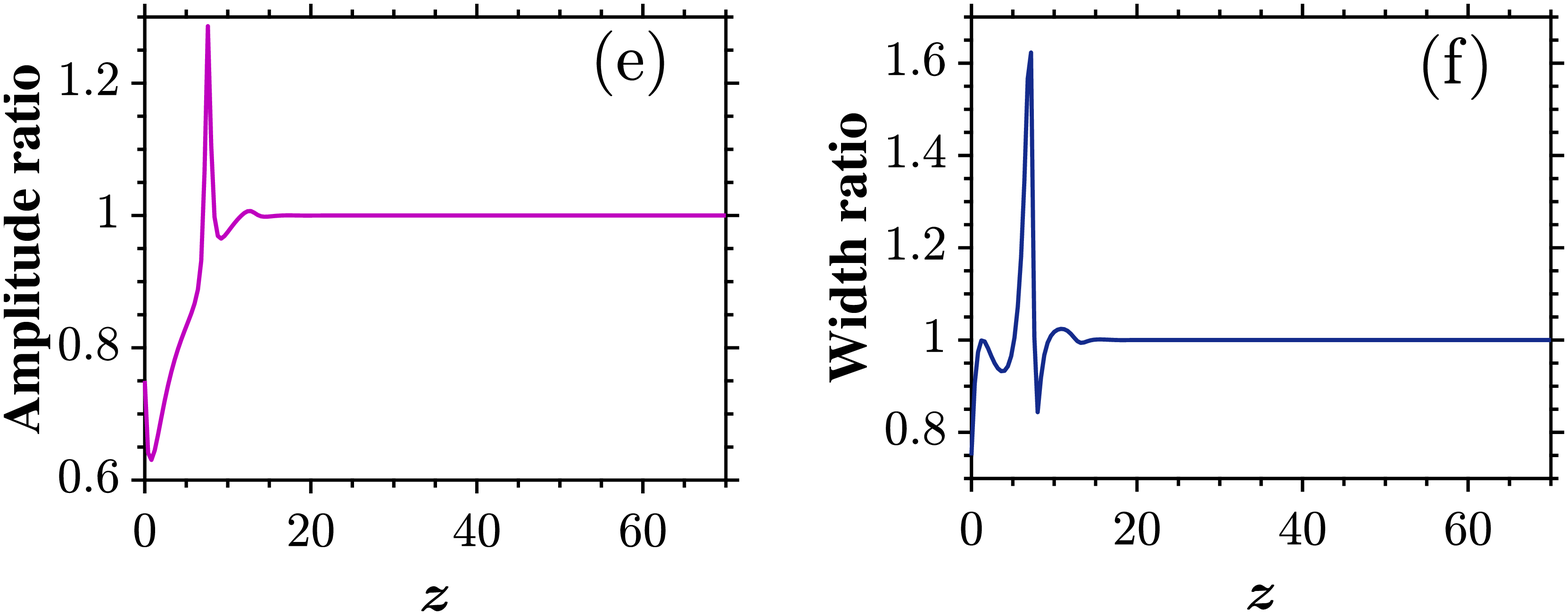}
\caption{\emph{ {\small Spatial evolution of amplitude ratio and
width ratio of input asymmetric gaussian pulses in the two cores.
(a-b): $u_{10}=u_{30}=1$, $v_{10}=0.75$, $v_{30}=1$;
  (c-d): $u_{10}=u_{30}=1$, $v_{10}=1$, $v_{30}=0.75$; and (e-f): $u_{10}=u_{30}=1$, $v_{10}=0.75$,
  $v_{30}=0.75$. The other system parameters are $k=0.01$, and  $d_r=-0.008$.}}}\label{fig3}
\end{figure}

 To launch the input pulses, we consider the asymmetric soliton like pulses (gaussian)
 that attain the asymmetric nature either in their amplitude or widths ($u_{i} \neq v_{i}$)
 expect the condition imposed on the temporal position as $u_{2}=-v_{2}$. Note that the other
parameters will  remain same in both cores as
($u_{40}=v_{40}=1$, $u_{50}=v_{50}=1$, $u_{60}=v_{60}=1$ and
$u_{2}=-v_{2}=1$).  As we have considered our model to be CQ-CGLEs
which are obtained as a perturbation to the standard NLSE, we assign
the parameters $p_i$, $q_i$, $c_i$ to be very small, viz., $ p=1-i0.6$, $
q=1-i0.046$, $ c=0.1+i0.0016$, $d_i=0.01$, $\gamma_r=0.001$ and
$\gamma_i=0$. Also, as mentioned in the description of
non-Lagrangian method, throughout the analysis numerical
computations will be done for the zero-average dispersion in the
fiber lasers.

We first study the role of TOD on the pulse evolution inside the
fiber cores as depicted in Fig. \ref{fig22}. As seen from the
figure, it is clear to observe that the width and amplitude in
each core fluctuate at the beginning of the longitudinal distance
prior to the stabilization (see Figs. \ref{fig22}(a) and
\ref{fig22}(b)). Once they settle themselves by stabilizing, the
pulses in both cores feature constant value with respect to
another neighboring core. The bottom panels in Fig. 1 reveals the
dynamics of amplitude and width ratios and it can be inferred
that the amplitude ratio and width ratio of waves propagating in
the two cores primarily oscillate before stabilizing to  unity in
their values after a certain distance.  One can further observe that we have also focused on
the roles of different input settings in the fiber lasers as shown
in Fig. \ref{fig3}. From this figure, it comes that when the initial asymmetry concerns only the
amplitudes \ref{fig3}(a-b), the amplitude and width ratios stabilize faster (the distance of fluctuations is
shorter) to unity and the peaks of fluctuations are lower. However, when
the initial asymmetry concerns the widths \ref{fig3}(c-d), the distance of fluctuations is larger. For total
asymmetry i.e asymmetry on initial amplitude and widths \ref{fig3}(e-f), the peaks of fluctuations of
amplitudes and widths is higher but the distance of fluctuations is shorter than in the second
case. One can then conclude that the initial asymmetry on amplitudes is more suitable for
shorter minimum length (i.e distance at which amplitudes and widths are superposed).

In addition to the above ramifications, we have also studied  the evolution of energies of waves
propagating inside each core. Their expressions are respectively given in the following relations
\begin{equation}\label{eq19}
\begin{split}
&E_1=\int _{-\infty }^{\infty }\! ( | f_1 |
) ^{2}{dt}=\sqrt {\frac{\pi}{2}}{{u_1}}^{2}{u_3}, \textrm{and}
 \end{split}
\end{equation}
\begin{equation}\label{eq20}
\begin{split}
& E_2=\int _{-\infty }^{\infty }\! ( | f_2 |) ^{2}{dt}=\sqrt
{\frac{\pi}{2}}{{v_1}}^{2}{v_3}.
 \end{split}
\end{equation}

\begin{figure}[t]
\centering
\includegraphics[width=0.3\linewidth]{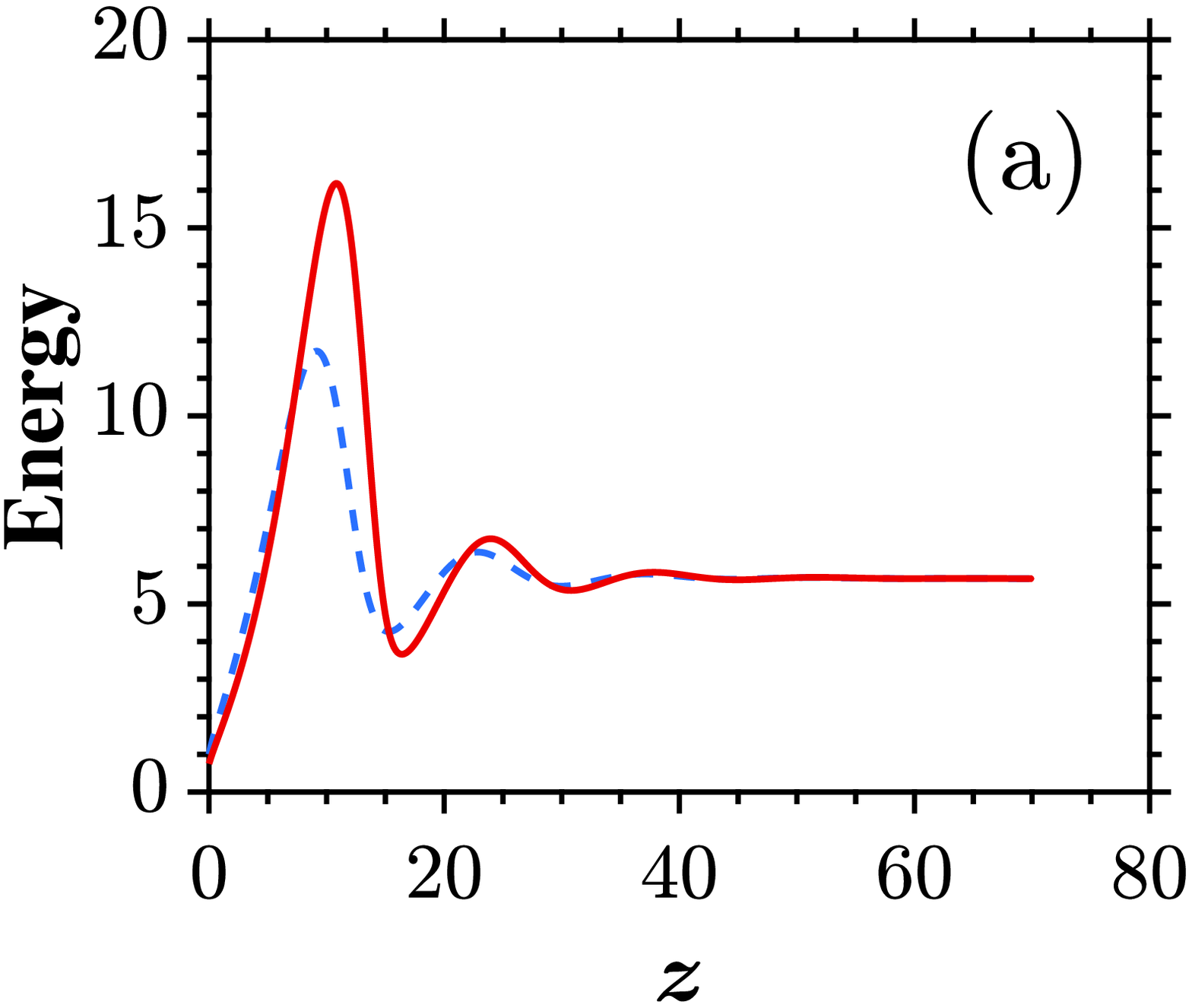}
    \includegraphics[width=0.3\linewidth]{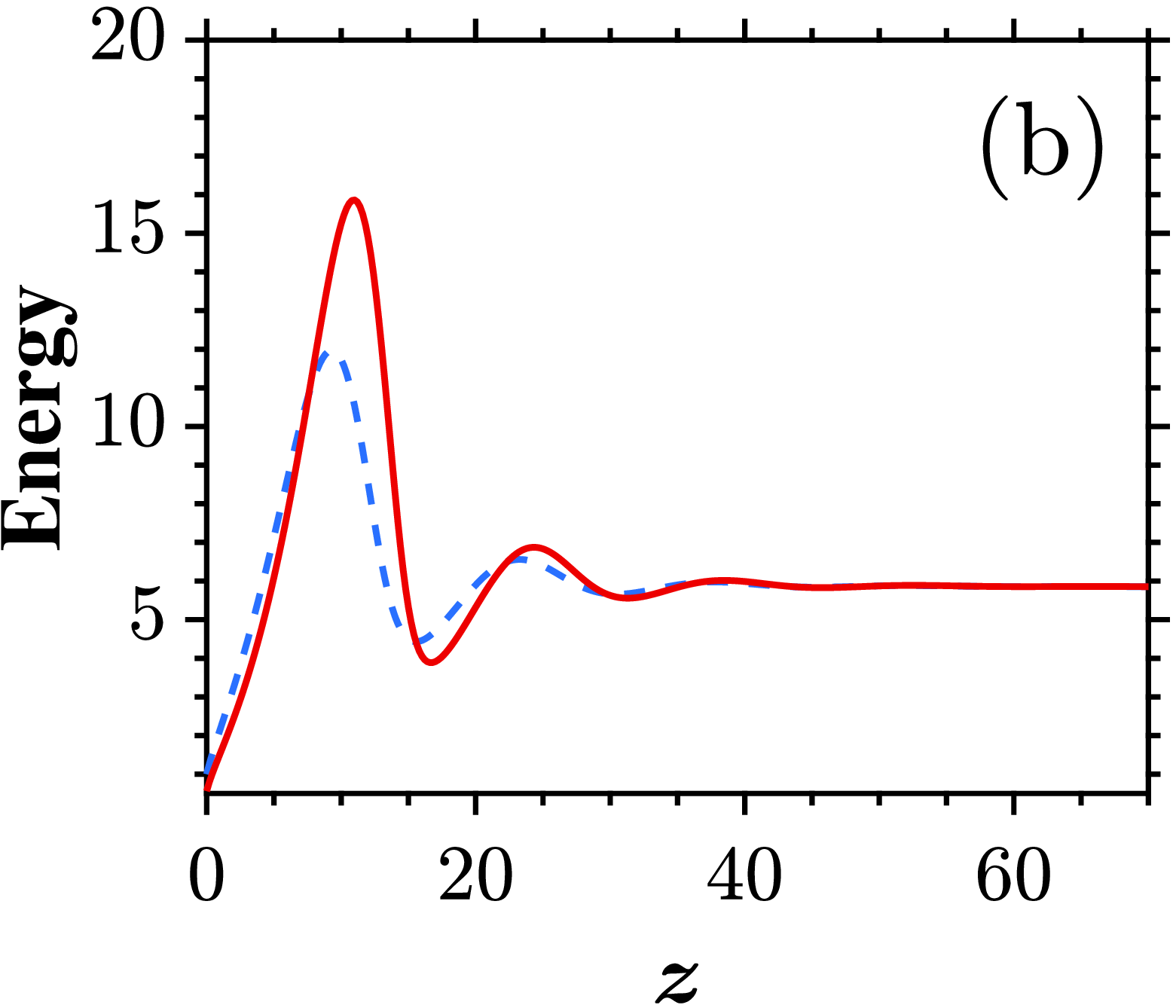}
    \includegraphics[width=0.3\linewidth]{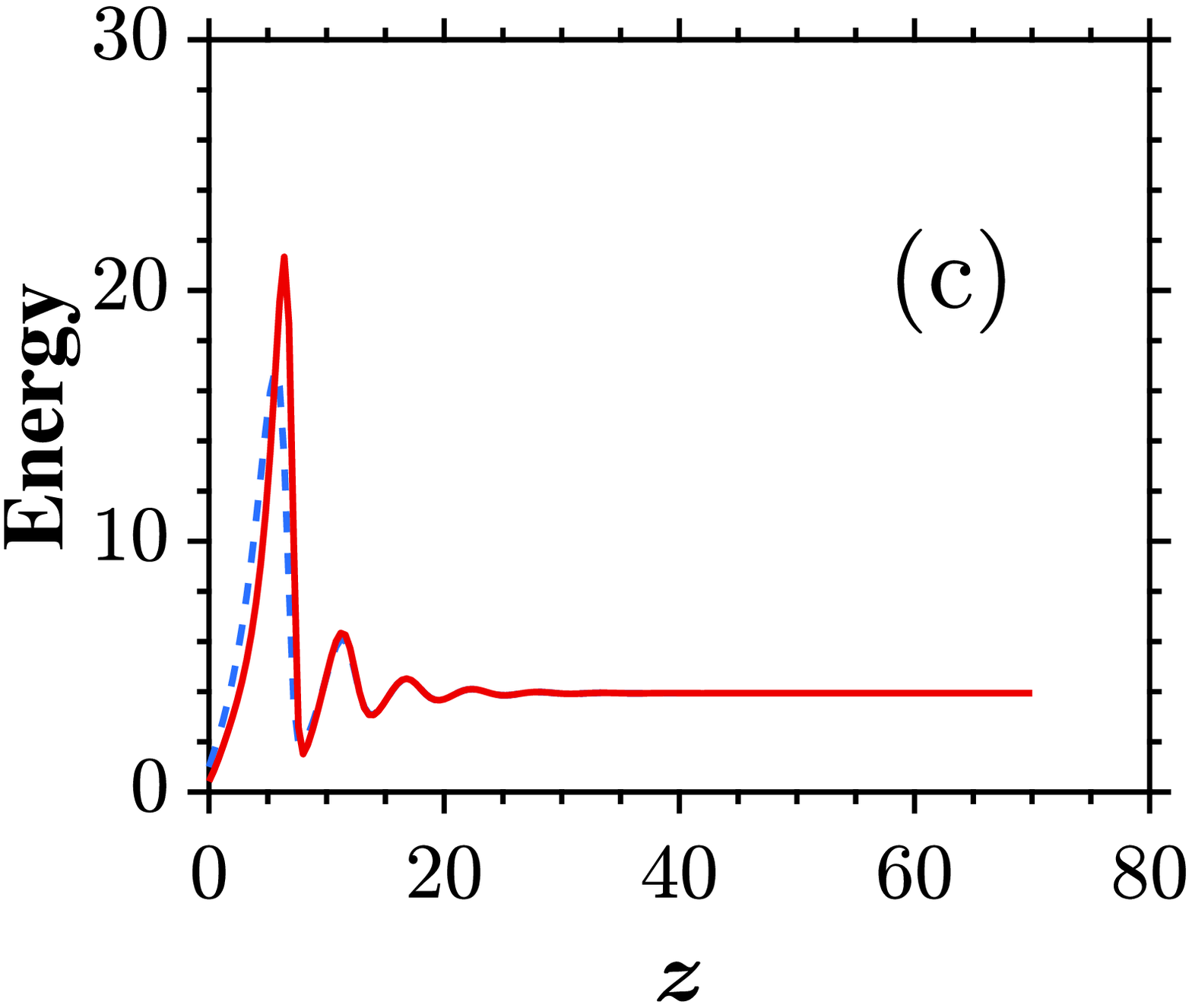}
\caption{\emph{ {\small Spatial evolution of energies of input
asymmetric gaussian pulses in the two cores of the fiber laser.
The blue curves correspond to the evolution in core 1 and the red
curves are meant for the evolution in core 2.  The system
parameters are (a): $u_{10}=u_{30}=1$, $v_{10}=0.75$, $v_{30}=1$;
  (b): $u_{10}=u_{30}=1$, $v_{10}=1$, $v_{30}=0.75$; and (c): $u_{10}=u_{30}=1$, $v_{10}=0.75$, $v_{30}=0.75$.
  In all plots, the other parameters are chosen such that $k=0.01$ and $d_r=-0.008$}}}\label{fig1}
\end{figure}
\begin{figure}[ht]
\centering
\includegraphics[width=0.3\linewidth]{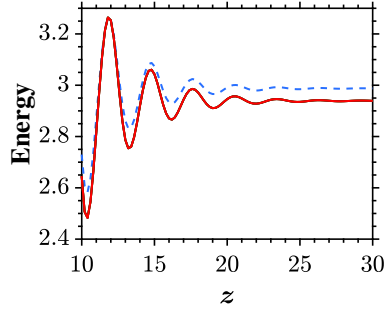}
    \caption{\emph{ {\small Impact of TOD on the spatial evolution of energies
    of input asymmetric gaussian pulses in the fiber laser. The blue curve
corresponds to $d_{r}=0.0$ and the red curve denotes $d_{r}=-0.1$. The
initial parameters values are $u_{10}=u_{30}=1$, $v_{10}=0.75$,
$v_{30}=1$, and $k=0.5$.}}}\label{fig11}
\end{figure}
Observations of Figs. \ref{fig22} and \ref{fig3} are confirmed in
Fig. \ref{fig1} which shows the variations of wave energies in
each core. One can observe from these figures that the cores
exchange their energies at the beginning of the propagation, the
one with the higher energy transfers it to the other core before
reaching an equilibrium state where the energies are superposed
(equal). Also, Fig. \ref{fig1} reveals that, the energies become equal
once the parameters (amplitudes and widths) get the same value.
Indeed in this case, both amplitude and width ratios are
equal to one. Besides, Fig. \ref{fig11} presents the effect of the
TOD on the energy in the first core at the equilibrium state. One can now observe that
in the presence of TOD ($d_r=-0.1$), the constant energy at the
equilibrium state is higher (red curve). In the following, we study the
influence of the coupling coefficient, initial parameters and the
TOD on the minimum distance after which the energy diagrams in
each core are superposed and determine numerically the values of
these distances in each case.
\subsection{Determining the minimum length of the fiber laser}
Following Eqs. \ref{eq19} and 18, the energy ratio
reads as
\begin{equation}\label{eq21}
\begin{split}
& E_{ratio}=\frac{E_2}{E_1}=\frac{{v_1}^{2}{v_3}}{{u_1}^{2}{u_3}}.
 \end{split}
\end{equation}
In order to obtain output
symmetric pulses,  the minimum length of the fiber laser is obtained numerically. This operation is performed
to determine the energy ratio closer to 1 ($E_{ratio}=1$) meaning that
  either the parameters are equal in both
cores i.e $u_{1}=v_{1}$ and $u_{3}=v_{3}$ or the energies of
pulses in the neighbouring cores are equal. Thus, we fix a
condition in the Runge-Kutta program as,
\begin{equation}\label{eq22}
|E_{ratio}-1|< \epsilon;
\end{equation}
where $\epsilon$ is an extremely small value, which takes
$\epsilon=10^{-6}$.
\subsubsection{Influence of inter-core coupling on the minimum length
of the fiber}
\begin{figure}[ht]
\centering
    \includegraphics[width=0.5\linewidth]{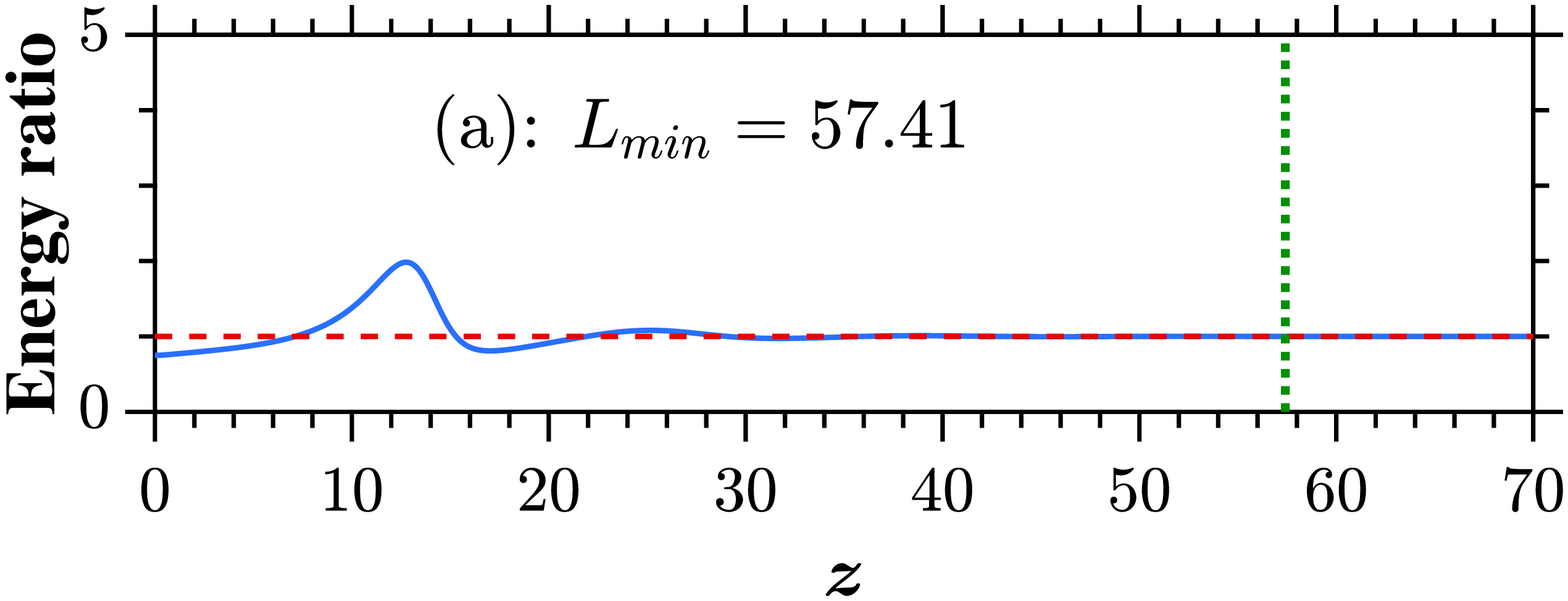}\\\includegraphics[width=0.5\linewidth]{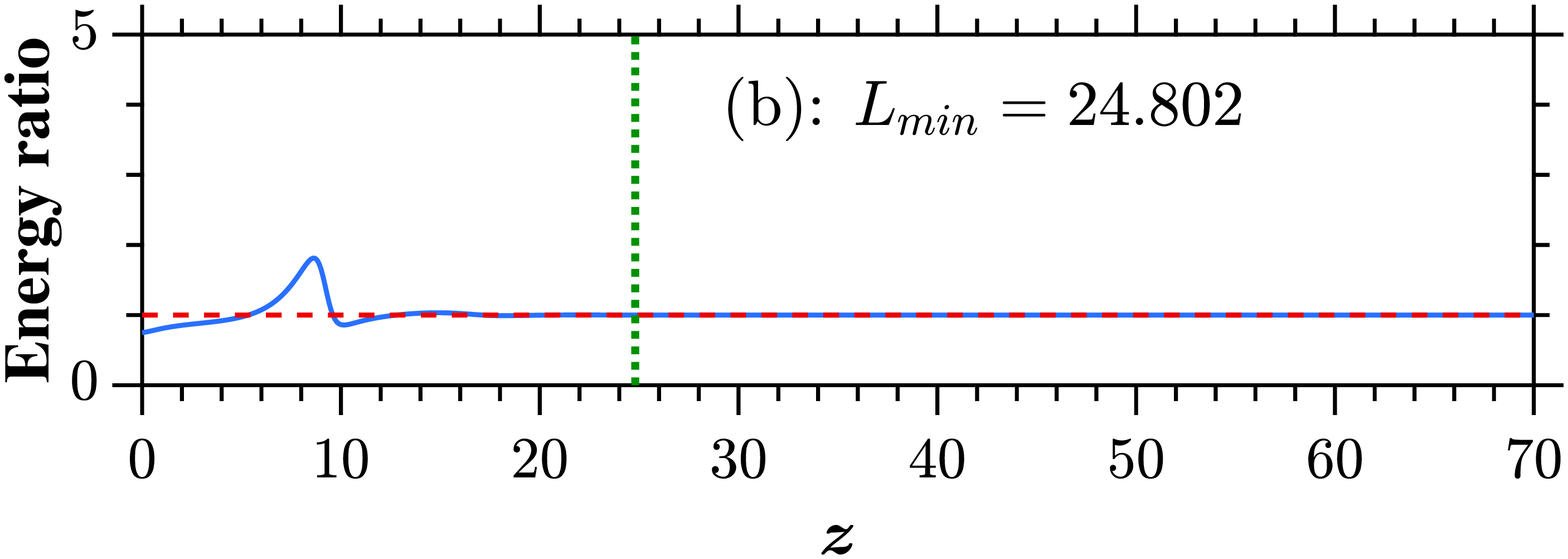}\\
    \includegraphics[width=0.5\linewidth]{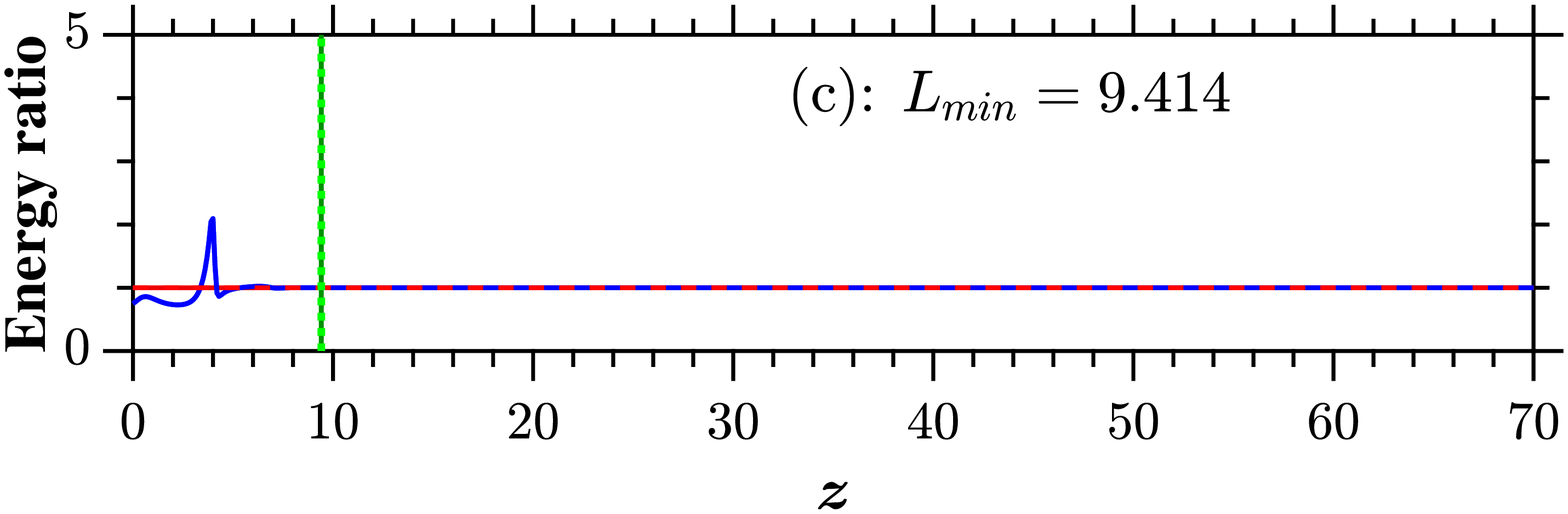}
\caption{Evolution of energy ratio for different values of the
coupling coefficient. (a): $k=0.01$,
  (b): $k=0.1$ and (c): $k=0.5$. The initial parameters values are
$u_{10}=u_{30}=1$, $v_{10}=0.75$, $v_{30}=1$, and $d_r=-0.008$}\label{fig5}
\end{figure}
Figure \ref{fig5} is plotted to investigate the spatial evolution of energy
ratio i.e the ratio between the energy of the pulse in core 2 over
the energy of the pulse in core 1 by adopting Eq. 19
for different values of the coupling coefficient. As seen in the figure, one can clearly notice that
the energy ratio stabilizes to 1  and more
faster on increasing the coupling constant. These values can be noted as follows, for $k=0.01$,
$L_{min}=57.410$ (Fig. \ref{fig5}(a)), for $k=0.1$,
$L_{min}=28.802$ (Fig. \ref{fig5}(b)) and for $k=0.5$,
$L_{min}=9.414$ (Fig. \ref{fig5}(c)). It is to be noted that the simulations are
made for input asymmetric parameters as follows; $u_{10}=1$,
$v_{10}=0.75$ and $u_{30}=v_{30}=1$. Let us remember the
dimensionless transformation \cite{djob2015study};
$x=\frac{X}{L_{D}}$, 
and $L_{D}=\frac{T_0^2}{|\beta_{2}|}$; X being the real distance,
$L_{D}$ the dispersion length, $T_0$ the width of the pulse and
$\beta_{2}$ the GVD coefficient (related to $p_r$ through the
relation $p_r=\frac{\beta_{2}}{2}$). We adopt the physical parameters
corresponding to standard nonlinear directional couplers, as
follows \cite{govind2015numerical}: $\beta_{2}=0.02 ps^2/m$, $T_0=50 fs$ at
$\lambda=1.5 \mu m$. From the latter, we estimate the minimum
length of the fiber respectively as follows, $28.705$, $14.401$ and
$4.707$ in meters.
\begin{figure}[ht]
\centering
    \includegraphics[width=0.5\linewidth]{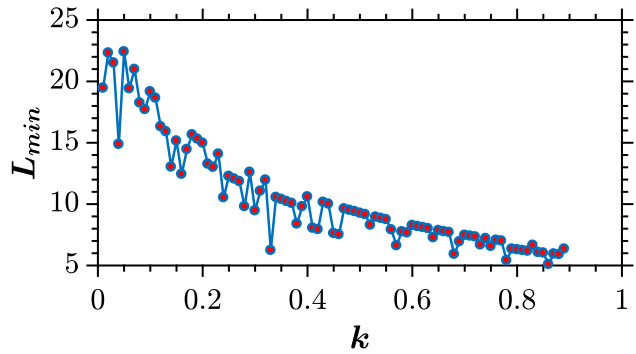}
\caption{Evolution of the minimum length versus the coupling
coefficient.  The system parameters are chosen such that $u_{10}=u_{30}=1$,
$v_{10}=0.75$, $v_{30}=1$ and
$d_r=-0.008$}\label{fig55}
\end{figure}

The previous observations pertaining to the influence of the
coupling constant on the minimum length are confirmed by Fig.
\ref{fig55} where $L_{min}$ is plotted versus k, the coupling
constant. This figure shows that $L_{min}$ decreases quasi
exponentially on increasing the value of $k$, even if there are
some fluctuations. For the strong coupling i.e, $k \rightarrow 1$,
the minimum length is quasi null ($L_{min} \rightarrow 0$).
\subsubsection{Influence of input amplitudes and widths on the minimum length
of the fiber laser}
\begin{figure}[ht]
\centering
    \includegraphics[width=0.5\linewidth]{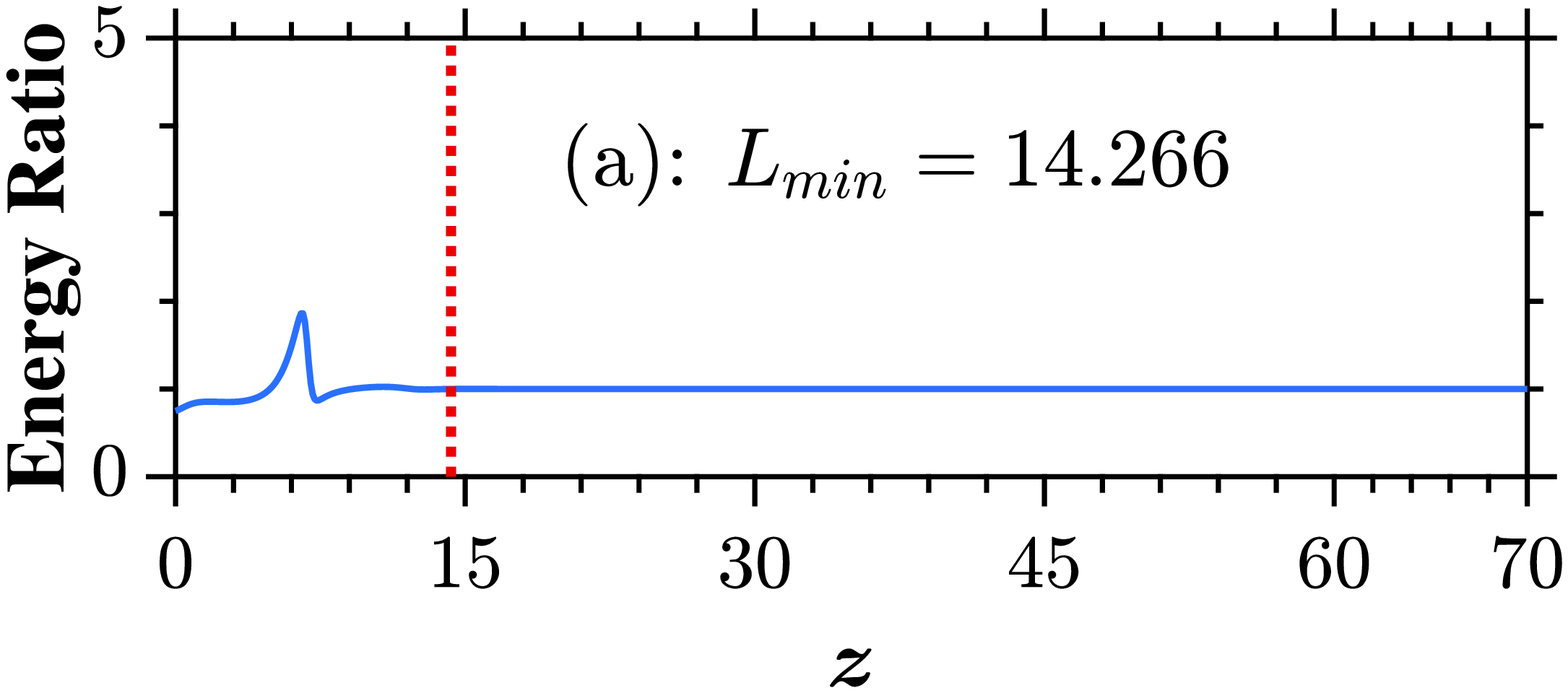}\\\includegraphics[width=0.5\linewidth]{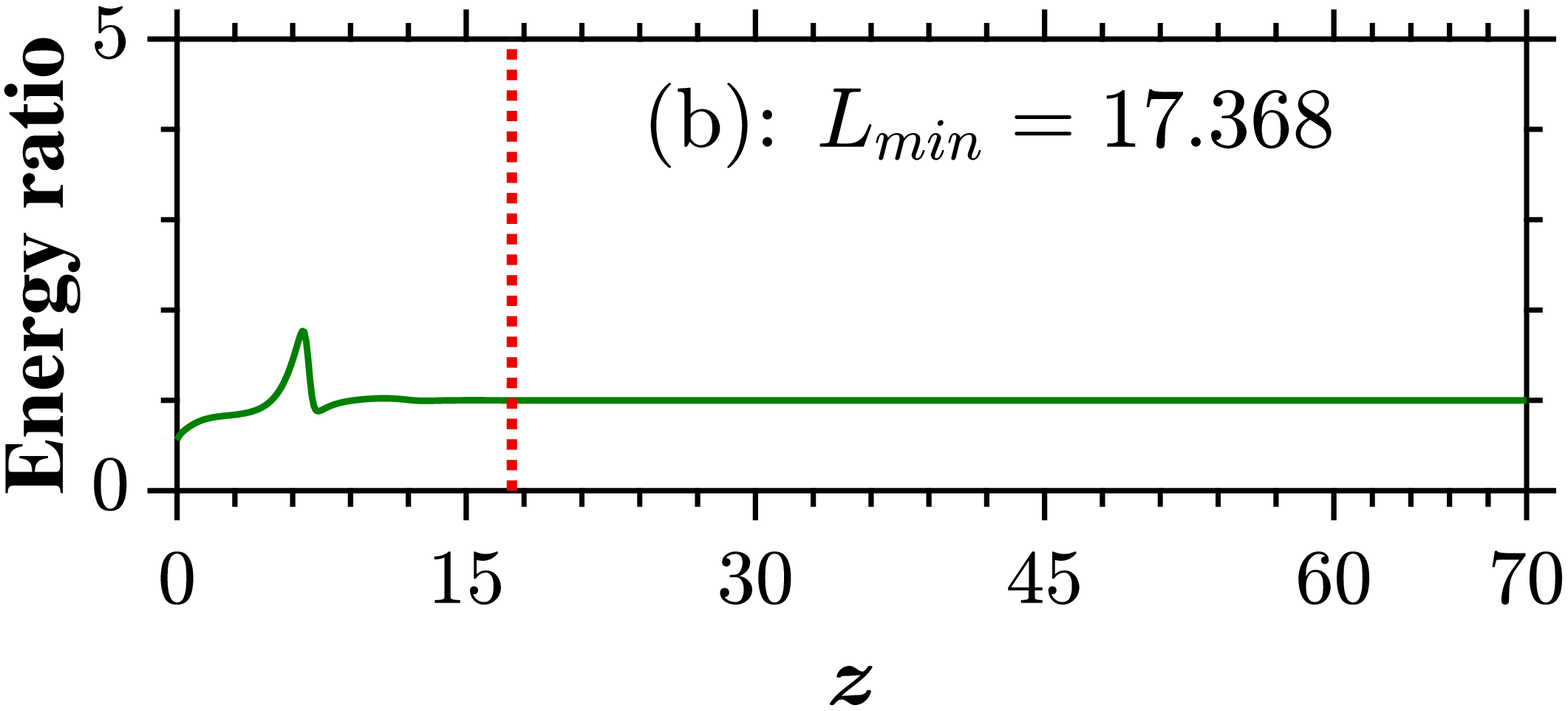}\\
    \includegraphics[width=0.5\linewidth]{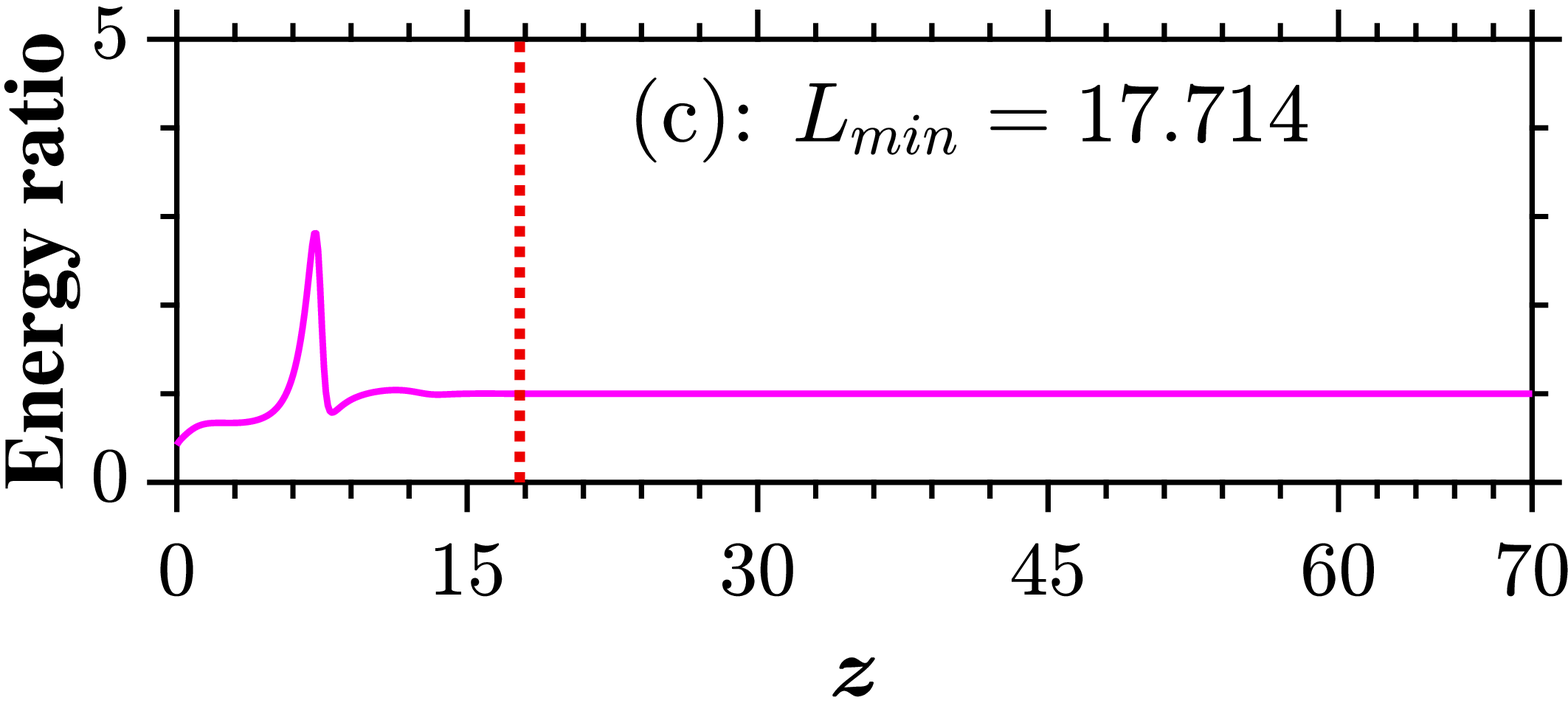}
\caption{Evolution of energy ratio for different values of
input parameters. (a): $u_{10}=1$, $v_{10}=0.75$,
$u_{30}=v_{30}=1$;
  (b): $u_{10}=v_{10}=1$, $u_{30}=1$, $v_{30}=0.75$; and (c): $u_{10}=1$, $v_{10}=0.75$, $u_{30}=1$, $v_{30}=0.75$. The other system parameters are
  $k=0.01$, and  $d_r=-0.008$}\label{fig6}
\end{figure}
Figure \ref{fig6} illustrates the spatial evolution of energy
ratio  for different values of the
input amplitudes and widths. One can find from the figure that the
minimum length is larger (as observed in Fig. \ref{fig6} (c)) corresponding to
full initial asymmetry viz.,  different input amplitudes and
different input widths. Indeed, we achieve the following minimum length as, $L_{min}=14.266$ when $u_{10}=1$,
$v_{10}=0.75$, $u_{30}=v_{30}=1$ (Fig. \ref{fig6}(a));
$L_{min}=17.368$ when $u_{10}=v_{10}=1$, $u_{30}=1$, $v_{30}=0.75$
(Fig. \ref{fig6}(b)) and $L_{min}=17.714$ when $u_{10}=1$,
$v_{10}=0.75$, $u_{30}=1$, $v_{30}=0.75$ (Fig. \ref{fig6}(c)).
From these ramifications, we estimate the minimum length of the fiber laser
respectively to: $7.133$, $8.684$ and $8.857$ in meters. Another
observation from the figure is that the maximum ratio is higher on
Fig. \ref{fig6}(c) compared to those of Fig. \ref{fig6}(a) and
Fig. \ref{fig6}(b). This is obvious since there is more
compensation when there are many input different parameters.
\subsubsection{Influence of the third order dispersion on the minimum length
of the fiber}
\begin{figure}[ht]
\centering
    \includegraphics[width=0.5\linewidth]{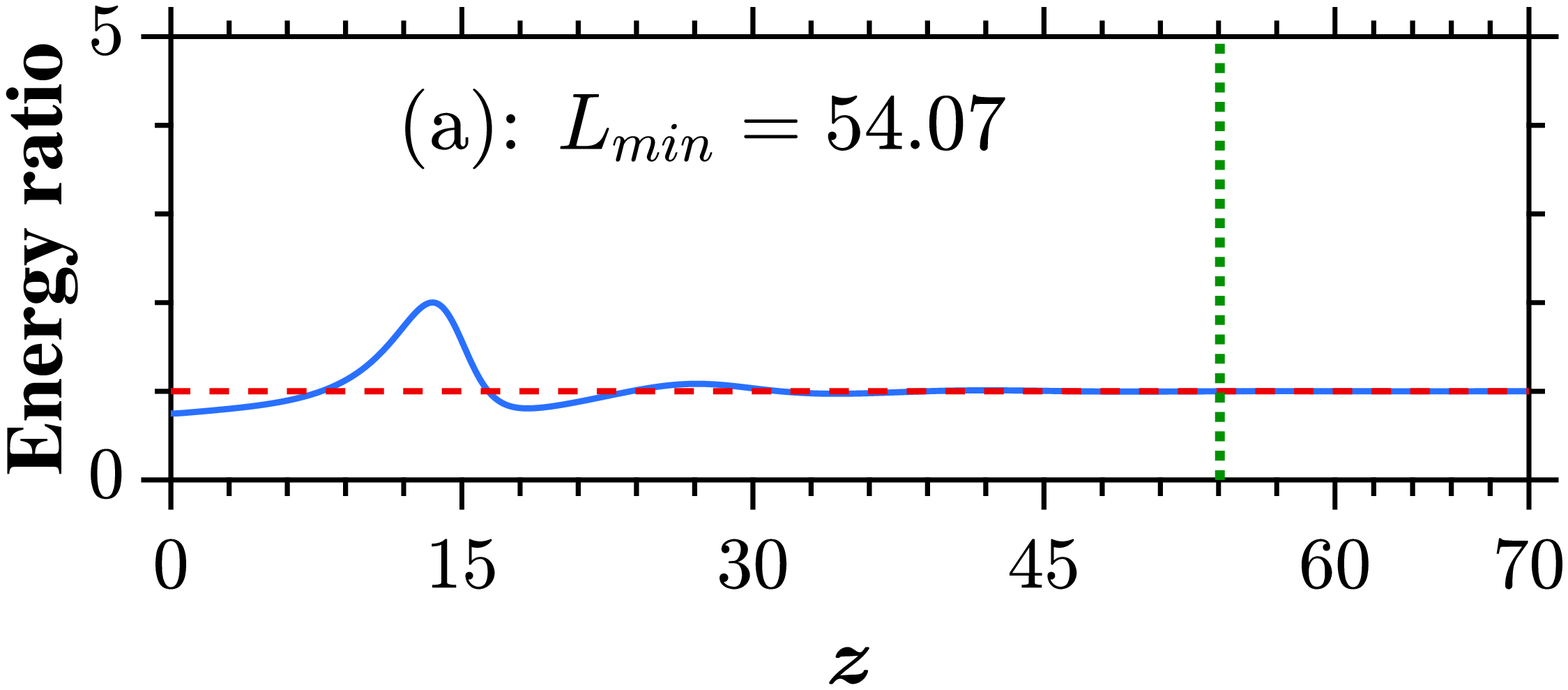}\\\includegraphics[width=0.5\linewidth]{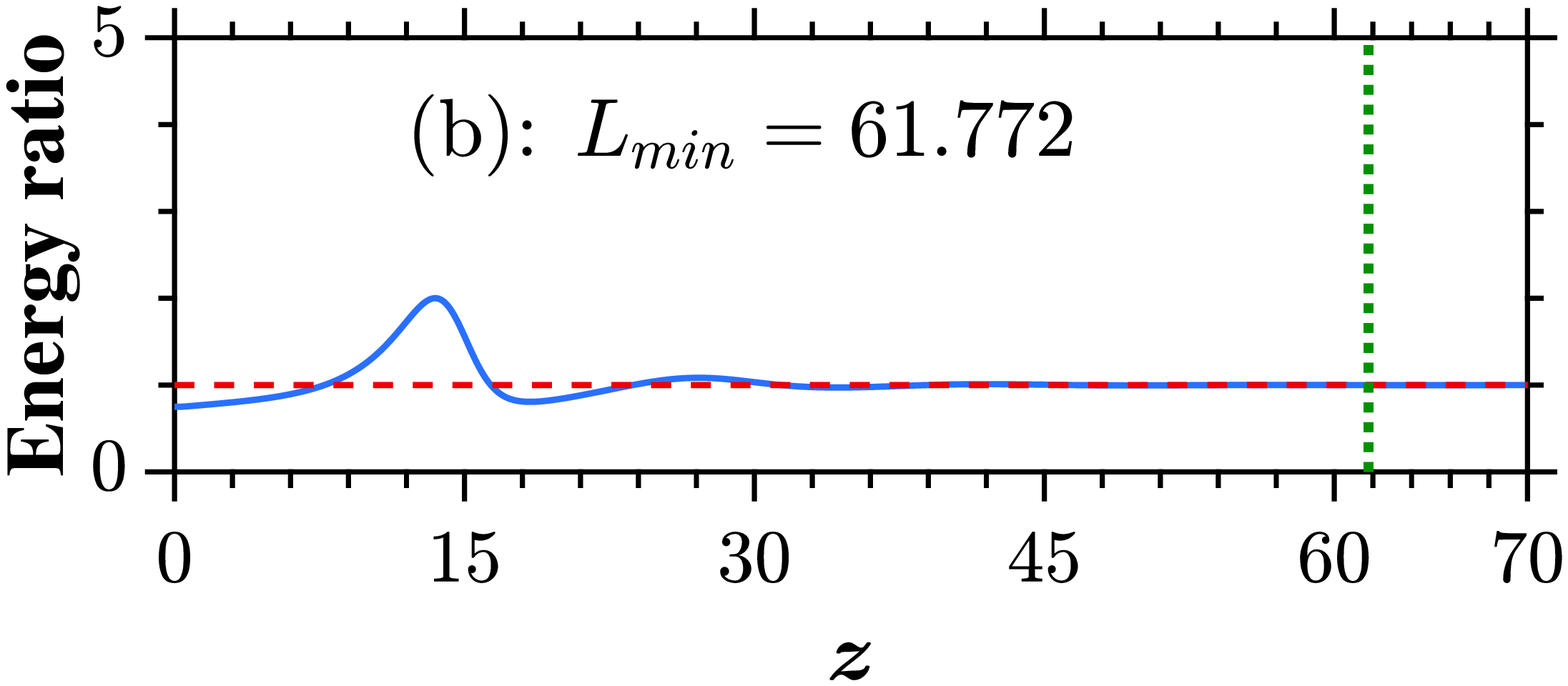}\\
    \includegraphics[width=0.5\linewidth]{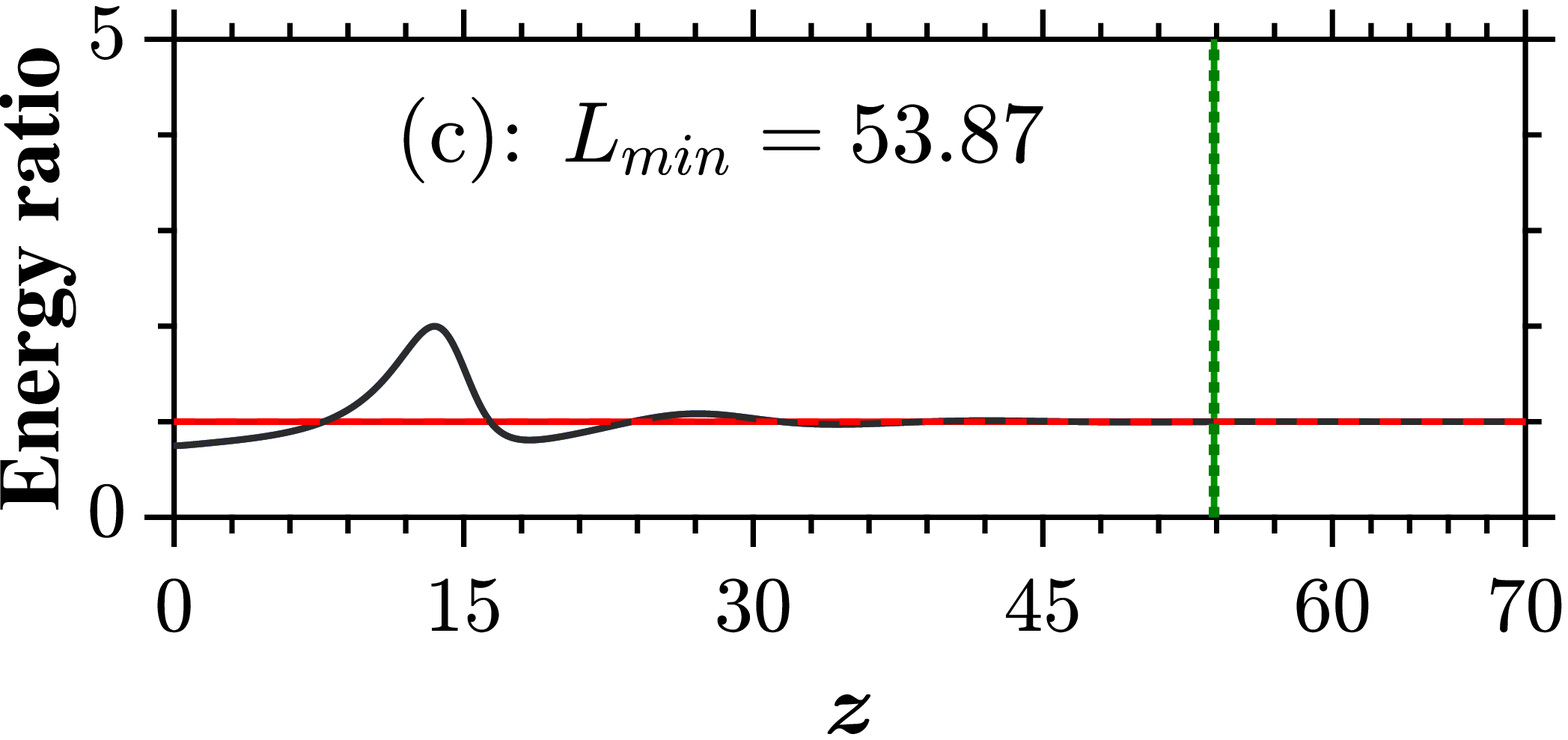}
\caption{Plots depicting evolution of energy ratio for different values of
 third order dispersion (TOD) coefficient. (a): $d_{r}=-0.008$;
  (b): $d_{r}=-0.01$; and (c): $d_{r}=-0.05$. The initial parameters values are $u_{10}=u_{30}=1$,
$v_{10}=0.75$, $v_{30}=1$, and
$k=0.01$.}\label{fig7}
\end{figure}
\begin{figure}[ht]
\centering
    \includegraphics[width=0.45\linewidth]{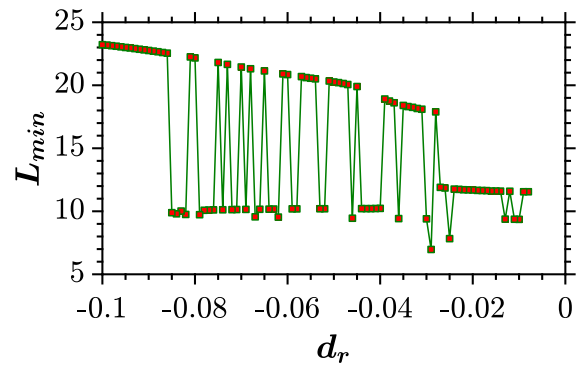}
\caption{Evolution of the minimum length versus the third order
dispersion (TOD) coefficient. The system parameters values are assigned
$u_{10}=u_{30}=1$, $v_{10}=0.75$, $v_{30}=1$, and  $k=0.5$.}\label{fig8}
\end{figure}

Figure \ref{fig7} depicting the spatial evolution of energy ratio
for different values of the third
order dispersion coefficient reveals that this TOD coefficient acts in
a strange manner. While increasing $d_{r}$, one was expecting the
minimum length to change in one way i.e may be increasing or
decreasing. Nevertheless, one observes that for $d_{r}=-0.008$,
$L_{min}$ is evaluated to $54.070$; for $d_{r}=-0.01$, $L_{min}$
is evaluated to $61.772$; while for $d_{r}=-0.05$, $L_{min}$ is
evaluated to $53.870$. Corresponding real values  are, respectively, $27.035$, $30.886$ and $26.935$ in meters. To explain these
observations, we have also plotted the evolution of the
minimum length versus the third order dispersion coefficient in Fig. \ref{fig8}. This
figure reveals a chaotic evolution of the minimum length beyond
two bifurcation points. Indeed, while varying the TOD coefficient
between $d_{r}=-0.025$ and $d_{r}=0.06$, we note that the minimum
length is quasi-constant (just few fluctuations) and relatively
lower. Before $d_{r}=-0.025$ and after $d_{r}=0.06$, the values of
the minimum lengths are unpredictable as the variations are in a random manner.
\newline
\section{Conclusions} \label{4}
We have done a systematic analysis pertaining to the dynamics of gaussian pulses in a system of fiber lasers by considering the third order parameter. From the results of the equations of motion of CVs parameters, we have
determined numerically the minimum distance of propagation
after which the pulses in two neighboring cores of a dual-core
fiber laser can remain in constant and equal energy. This distance
corresponds to the minimum length of the fiber for which
asymmetric input pulses should result in symmetric output pulses. We have successfully obtained these values numerically by the means of
the Runge Kutta algorithm following a non-Lagrangian approach. The analysis of the results show that
the minimum length of the fiber is strongly influenced by the
coupling, the initial asymmetry and the TOD. We have also found that strong coupling and total initial asymmetry favors the shorter minimum length of the fiber and we hope that these outcomes are
beneficial for safe propagation of the neighboring pulses in a
nonlinear dual-core fiber lasers. A remarkable aspect is noted
when stressing the impact of TOD influence, namely, this parameter increases the common
width and energy at the equilibrium state. We have also discovered a chaotic pattern
beyond some bifurcation values while varying the TOD
coefficient. 
\appendix
\section{Equations of motions}
\begin{equation} \label{eq20}
\begin{split}
\dot{u_1}=& -\frac {3}{2}\,[\frac {\sqrt {2}}{  ( {{\it u_3}}^{2}+{{\it v_3}}^{2}
    ) ^{\frac {5}{2}}{{\it u_3}}^{2}} [ \frac {4}{3}\,{\it v_1}\, ( -\frac {3}{4}\,{{
        \it u_3}}^{4}+ ( ({\it u_2}^2-{\it v_2})^2-{\it v_3}^2  ) {{\it
        u_3}}^{2}-\frac {1}{4}\,{{\it v_3}}^{4}
) k{{\it u_3}}^{2}{\it v_3}\,{{\rm e}^{-{\frac { ( {\it u_2}-
                {\it v_2}  ) ^{2}}{{{\it u_3}}^{2}+{{\it v_3}}^{2}}}}} \\
&+ (
(  ( -\frac {1}{3}\,d_i{{\it u_5}}^{3}+{\it d_r}\,{\it u_4}\,{\it u_5}-\frac {1}{3}
\,{{\it u_5}}^{2}p_i+\frac {1}{3}\,{\it u_4}\,{\it p_r}-\frac{\it
    \gamma_r}{3}  ) {{ \it u_3}}^{2}-2\,d_i{\it u_5}-\frac
{2}{3}\,p_i  ) \sqrt {2}\\
& +{\frac {4\,{{\it u_1 }}^{2}{{\it
                u_3}}^{2}}{27} ( {{\it u_1}}^{2}\sqrt {6}{\it c_i}+{ \frac
        {45\,{\it qi}}{16}} ) }  ) {\it u_1}\, ( {{\it u_3 }}^{2}+{{\it
        v_3}}^{2} ) ^{\frac {5}{2}} ] ]\\
\dot{u_2}=& \frac{3}{16}\,[\frac {\sqrt {2}}{ ( {{\it u_3}}^{2}+{{\it v_3}}^{2}
    ) ^{\frac{3}{2}}{{\it u_3}}^{2}{\it u_1}} ( -{\frac {32\,k{\it
            v_1}\, {\it v_3}\,{{\it u_3}}^{4} ( {\it u_2}-{\it v_2} )
    }{3}{{\rm e} ^{-{\frac { ( {\it u_2}-{\it v_2}) ^{2}}{{{\it
                            u_3}}^{2}+{{ \it v_3}}^{2}}}}}}+ ( d_i{{\it u_3}}^{6}{{\it
        u_4}}^{3}+2\,{\it u_4}\,
( 2\,d_i{{\it u_5}}^{2}\\
&+{\it d_r}\,{\it u_4}+\frac{4}{3}\,p_i{\it u_5}) {
    {\it u_3}}^{4}+ ( 4\,d_i{\it u_4}+8\,{\it u_5}\, ( {\it u_5}\,{
    \it d_r}+\frac{2}{3}\,{\it p_r}) ) {{\it u_3}}^{2}+8\,{\it d_r} )
{\it u_1}\,\sqrt {2} ( {{\it u_3}}^{2}+{{\it v_3}}^{2}
) ^{{3}{2}}) ]\\
\dot{u_3}=& \frac{3}{4}\,{\frac {\sqrt {2}}{ ( {{\it u_3}}^{2}+{{\it v_3}}^{2} )
        ^{\frac{5}216}}{\it u_3}\,{\it u_1}} [ \frac{16}{3}\,{\it v_1}\, (
-\frac{1}{4}\,{{\it u_3}}^{4}+{{\it u_3}}^{2} ( {\it u_2}-{\it
    v_2} ) ^{2}+\frac{1}{4}\,{{ \it v_3}}^{4} ) k{{\it u_3}}^{2}{\it
    v_3}\,{{\rm e}^{-{\frac {
                ( {\it u_2}-{\it v_2} ) ^{2}}{{{\it u_3}}^{2}+{{\it v_3}}^{2}}
}}}+ ( ( {{\it u_4}}^{2} ( d_i{\it u_5}+\frac{\it p_i}{3} ) {{\it
        u_3}}^{4}\\
&+4\,{\it u_4}\, ( {\it u_5}\,{\it d_r}+\frac{\it p_r}{3} )
{{ \it u_3}}^{2}-4\,d{\it u_5}-\frac{4}{3}\,p_i ) \sqrt {2}+{\frac
    {4\,{{\it u_1 }}^{2}( {{\it u_1}}^{2}\sqrt {6}{\it
            c_i}+\frac{9}{4}\,{\it q_i} ) {{ \it u_3}}^{2}}{27}} ) {\it u_1}\,
( {{\it u_3}}^{2}+{{\it v_3}} ^{2} ) ^{\frac{5}{2}} ]\\
\dot{u_4}=& -3\,[\frac {\sqrt {2}}{ ( {{\it u_3}}^{2}+{{\it
            v_3}}^{2} ) ^ {\frac{5}{2}}{{\it u_3}}^{4}{\it u_1}} (
-\frac{16}{3}\,{\it v_1}\, (  ( \frac{1}{2}\,{{\it v_3}}^{2}+ (
{\it u_2}-{\it v_2}) ^{2}) {{ \it u_3}}^{2}+\frac{1}{2}\,{{\it
        v_3}}^{4} ) k{{\it u_3}}^{2}{\it v_3}\,{ {\rm e}^{-{\frac { ( {\it
                    u_2}-{\it v_2}
                ) ^{2}}{{{\it u_3}}^{ 2}+{{\it v_3}}^{2}}}}}\\
&+ ( ( {{\it u_4}}^{2} (
{\it u_5}\, {\it dr}+\frac{\it p_r}{3} ) {{\it u_3}}^{4}-4\,(
d{\it u_5}+\frac{p_i}{3} ) {\it u_4}\,{{\it u_3}}^{2}-4\,{\it
    u_5}\,{\it d_r}-\frac{4}{3}\,{\it p_r} ) \sqrt {2}+{\frac
    {4\,{{\it u_1}}^{2} ( {{\it u_1}}^{2} \sqrt {6}{\it
            c_r}+\frac{9}{4}\,{\it q_r}) {{\it
                u_3}}^{2}}{27}})\\
& { \it u_1}\, ( {{\it u_3}}^{2}+{{\it v_3}}^{2} )
^{5/2}) ]\\
\dot{u_5}=& \frac{3}{16}\,[\frac {\sqrt {2}}{ ( {{\it
            u_3}}^{2}+{{\it v_3}}^{2} ) ^{\frac{3}{2}}{{\it u_3}}^{4}{\it
        u_1}} ( -{\frac {32\,{{\it u_3}}^ {6}{\it u_4}\,k{\it v_1}\,{\it
            v_3}\, ( {\it u_2}-{\it v_2} ) }{ 3}{{\rm e}^{-{\frac { ( {\it
                        u_2}-{\it v_2} ) ^{2}}{{{\it u_3} }^{2}+{{\it v_3}}^{2}}}}}}+ (
d_i{{\it u_3}}^{4}{{\it u_4}}^{2}+
( 4\,d_i{{\it u_5}}^{2}+\frac{8}{3}\,p_i{\it u_5}) {{\it
        u_3}}^{2}\\
&+4\,d_i
)  ( {{\it u_3}}^{4}{{\it u_4}}^{2}+4) {\it u_1}\,
\sqrt {2} ( {{\it u_3}}^{2}+{{\it v_3}}^{2} ) ^{\frac{3}{2}} ) ]\\
\dot{u_6}=& \frac{3}{16}\,[\frac {\sqrt {2}}{ ( {{\it
            u_3}}^{2}+{{\it v_3}}^{2}
    ) ^{\frac{5}{2}}{{\it u_3}}^{2}{\it u_1}} ( \\
&-{\frac {32\,{\it v_1}\,k
        ( {\it u_5}\, ( {\it u_2}-{\it v_2} ) {{\it u_3}}^{4}+ ( (
        \frac{1}{2}+{\it u_5}\, ( {\it u_2}-{\it v_2} ) ) {{\it v_3}}^{2}+
        ( {\it u_2}-{\it v_2}) ^{2}
        ) {{\it u_3}}^{2}+\frac{1}{2}\,{{\it v_3}}^{4} ) {{\it u_3}}^{2}{
            \it v_3}}{3}{{\rm e}^{-{\frac { ( {\it u_2}-{\it v_2}) ^{2}}{
                    {{\it
                            u_3}}^{2}+{{\it v_3}}^{2}}}}}}\\
&+ (  ( d_i{{\it u_3}}^{6}{{ \it
        u_4}}^{3}{\it u_5}+2\,{\it u_5}\,{\it u_4}\, ( 2\,d_i{{\it
        u_5}}^{2} +{\it d_r}\,{\it u_4}+\frac{4}{3}\,p_i{\it u_5} ) {{\it
        u_3}}^{4}+ ( -4\, d_i{\it u_4}\,{\it u_5}+\frac{8}{3}\,{\it
    \gamma_i}+\frac{8}{3}\,{\it p_r}\,{{\it
        u_5}}^{2}-\frac{8}{3}\,p_i{\it
    u_4}\\
&+\frac{16}{3}\,{\it d_r}\,{{\it u_5}}^{3} ) {{\it u_3}}^{2}-\frac{16}{3}\,{\it p_r}-8\,{\it u_5}\,{\it d_r} ) \sqrt {2}+{\frac {32\,{{\it
                u_1}}^{2}{{\it u_3}}^{2}}{27} ( {{\it u_1}}^{2}\sqrt {6}{\it
        c_r}+{ \frac {45\,{\it q_r}}{16}} ) } ) {\it u_1}\, ( {{\it u_3
}}^{2}+{{\it v_3}}^{2} ) ^{\frac{5}{2}} ) ] ,
\end{split}
\end{equation}
and:
\begin{equation} \label{eq21}
\begin{split}
\dot{v_1}=& -\frac{2}{9}\,[\frac {\sqrt {2}}{ ( {{\it u_3}}^{2}+{{\it v_3}}^{2}
    ) ^{\frac{5}{2}}{{\it v_3}}^{2}} ( 9\,{{\it v_3}}^{2}{\it u_1}\,
( -\frac{3}{4}\,{{\it v_3}}^{4}+( -{\it u_3}+{\it u_2}-{\it v_2} )
( {\it u_3}+{\it u_2}-{\it v_2} ) {{\it v_3}}^{2}-\frac{1}{4}
\,{{\it u_3}}^{4}) k{\it u_3}\,{{\rm e}^{-{\frac { ( {\it u_2
                }-{\it v_2} )
                ^{2}}{{{\it u_3}}^{2}+{{\it v_3}}^{2}}}}}\\
&+ ( {{ \it u_3}}^{2}+{{\it
        v_3}}^{2} ) ^{\frac{5}{2}} (  ( ( -\frac{9}{4} \,d_i{{\it
        v_5}}^{3}+{\frac {27\,{\it d_r}\,{\it v_4}\,{\it
            v_5}}{4}}-\frac{9}{4}\,p_i {{\it v_5}}^{2}+\frac{9}{4}\,{\it
    p_r}\,{\it v_4}-\frac{9}{4}\,{\it \gamma_r} ) {{ \it
        v_3}}^{2}-{\frac {27\,d_i{\it v_5}}{2}}-\frac{9}{2}\,p_i ) \sqrt
{2}\\
&+{{ \it v_1}}^{2}{{\it v_3}}^{2} ( {{\it v_1}}^{2}\sqrt {6}{\it
    c_i}+{ \frac {45\,{\it q_i}}{16}} ) ) {\it v_1} ) ]\\
\dot{v_2}=& \frac{3}{16}\,[\frac {\sqrt {2}}{( {{\it u_3}}^{2}+{{\it v_3}}^{2}
    ) ^{\frac{3}{2}}{{\it v_3}}^{2}{\it v_1}} ( {\frac {32\,k{{\it v_3}}^
        {4}{\it u_3}\,{\it u_1}\, ( {\it u_2}-{\it v_2} ) }{3}{{\rm e}^
        {-{\frac {( {\it u_2}-{\it v_2}) ^{2}}{{{\it u_3}}^{2}+{{ \it
                            v_3}}^{2}}}}}}+ ( {{\it u_3}}^{2}+{{\it v_3}}^{2} ) ^{\frac{3}{2}}
\sqrt {2}{\it v_1}\, ( d_i{{\it v_3}}^{6}{{\it v_4}}^{3}\\
&+2\, ( 2
\,d_i{{\it v_5}}^{2}+{\it d_r}\,{\it v_4}+\frac{4}{3}\,p_i{\it
    v_5} ) {\it v_4}\, {{\it v_3}}^{4}+ ( 4\,d_i{\it v_4}+8\, ( {\it
    v_5}\,{\it dr}+\frac{2}{3} \,{\it p_r}) {\it v_5} ) {{\it
        v_3}}^{2}+8\,{\it d_r}
)  ) ]\\
\dot{v_3}=& \frac{1}{9}\,[\frac {\sqrt {2}}{ ( {{\it u_3}}^{2}+{{\it v_3}}^{2} )
    ^{\frac{5}{2}}{\it v_1}\,{\it v_3}} ( 36\,{{\it v_3}}^{2}{\it u_1}\,k\\
&
( -\frac{1}{4}\,{{\it v_3}}^{4}+{{\it v_3}}^{2} ( {\it u_2}-{\it
    v_2} ) ^ {2}+\frac{1}{4}\,{{\it u_3}}^{4} ) {\it u_3}\,{{\rm
        e}^{-{\frac { ( { \it u_2}-{\it v_2} ) ^{2}}{{{\it u_3}}^{2}+{{\it
                        v_3}}^{2}}}}}+
( {{\it u_3}}^{2}+{{\it v_3}}^{2} ) ^{\frac{5}{2}} (  ( {
    \frac { ( 27\,d_i{\it v_5}+9\,p_i ) {{\it v_4}}^{2}{{\it v_3}}^{4}
    }{4}}\\
&+27\, ( {\it v_5}\,{\it d_r}+\frac{p_r}{3} ) {\it v_4}\,{{ \it
        v_3}}^{2}-27\,d_i{\it v_5}-9\,p_i ) \sqrt {2}+{{\it v_1}}^{2}{{\it
        v_3}}^{2} ( {{\it v_1}}^{2}\sqrt {6}{\it c_i}+\frac{9}{4}\,{\it
    q_i} ) ) {\it v_1} ) ]\\
\dot{v_4}=& -\frac{4}{9}\,[\frac {\sqrt {2}}{ ( {{\it
            u_3}}^{2}+{{\it v_3}}^{2}
    ) ^{\frac{5}{2}}{{\it v_3}}^{4}{\it v_1}} ( -36\,{{\it v_3}}^{2}
(  ( \frac{1}{2}\,{{\it u_3}}^{2}+ ( {\it u_2}-{\it v_2} )
^{2} ) {{\it v_3}}^{2}+\frac{1}{2}\,{{\it u_3}}^{4}) {\it u_1}\,k{
    \it u_3}\,{{\rm e}^{-{\frac { ( {\it u_2}-{\it v_2} ) ^{2}}{{{ \it
                        u_3}}^{2}+{{\it v_3}}^{2}}}}}\\
&+ ( {{\it u_3}}^{2}+{{\it v_3}}^{2}
) ^{\frac{5}{2}} (  ( {\frac { ( 27\,{\it v_5}\,{\it d_r}+
        9\,{\it p_r} ) {{\it v_4}}^{2}{{\it v_3}}^{4}}{4}}-27\, ( d_i{ \it
    v_5}+\frac{p_i}{3} ) {\it v_4}\,{{\it v_3}}^{2}-27\,{\it
    v_5}\,{\it
    d_r}-9 \,{\it p_r} ) \sqrt {2}\\
&+{{\it v_1}}^{2}{{\it v_3}}^{2} ( {{
        \it v_1}}^{2}\sqrt {6}{\it c_r}+\frac{9}{4}\,{\it q_r} ) ) {\it
    v_1}
) ]\\
\dot{v_5}=& 2\,[\frac {\sqrt {2}}{( {{\it u_3}}^{2}+{{\it
            v_3}}^{2} ) ^{ \frac{3}{2}}{{\it v_3}}^{4}{\it v_1}} ( {\frac {3\,
        ( {{\it u_3}}^{2}+{ {\it v_3}}^{2} ) ^{\frac{3}{2}} ( d_i{{\it
                v_3}}^{4}{{\it v_4}}^{2}+ ( 4\,d_i{{\it
                v_5}}^{2}+\frac{8}{3}\,p_i{\it v_5} ) {{\it v_3}}^{2}+4\,d_i )
        \sqrt {2}( {{\it v_3}}^{4}{{\it v_4}}^{2}+4 ) {\it
            v_1}}{32}}\\
&+{{\it v_3}}^{6}{\it v_4}\,k{\it u_3}\,{\it u_1}\, ( {\it u_2
}-{\it v_2} ) {{\rm e}^{-{\frac { ( {\it u_2}-{\it v_2}
                ) ^{2}}{{{\it u_3}}^{2}+{{\it v_3}}^{2}}}}} ) ]\\
\dot{v_6}=& \frac{2}{9}\,[\frac {\sqrt {2}}{ ( {{\it
            u_3}}^{2}+{{\it v_3}}^{2} ) ^{\frac{5}{2}}{{\it v_3}}^{2}{\it
        v_1}} ( -9\,{{\it v_3}}^{2} ( -{\it v_5}\, ( {\it u_2}-{\it v_2} )
{{\it v_3}}^{4}+ ( ( \frac{1}{2}+ ( {\it v_2}-{\it u_2} ) {\it
    v_5}) {{\it u_3}}^{2}+
( {\it u_2}-{\it v_2} ) ^{2} ) {{\it v_3}}^{2}\\
&+\frac{1}{2}\,{{
        \it u_3}}^{4} ) {\it u_1}\,k{\it u_3}\,{{\rm e}^{-{\frac { ( { \it
                    u_2}-{\it v_2} ) ^{2}}{{{\it u_3}}^{2}+{{\it v_3}}^{2}}}}}+ ( (
{\frac {27\,d_i{{\it v_3}}^{6}{{\it v_4}}^{3}{\it v_5}}{32}
}+{\frac {27\,{\it v_5}\, ( 2\,d_i{{\it v_5}}^{2}+{\it d_r}\,{\it
            v_4}+ 4/3\,p_i{\it v_5} ) {\it v_4}\,{{\it v_3}}^{4}}{16}}+ (
-{\frac {27\,d_i{\it v_4}\,{\it v_5}}{8}}+\frac{9}{4}\,{\it
    \gamma_i}\\
&+\frac{9}{4}\,{{\it v_5}}^{2}{ \it
    p_r}-\frac{9}{4}\,p_i{\it v_4}+\frac{9}{2}\,{{\it v_5}}^{3}{\it
    d_r} ) {{\it v_3}}^ {2}-\frac{9}{2}\,{\it p_r}-{\frac {27\,{\it
            v_5}\,{\it d_r}}{4}}) \sqrt {2 }+{{\it v_1}}^{2}{{\it v_3}}^{2} (
{{\it v_1}}^{2}\sqrt {6}{\it c_r}+ {\frac {45\,{\it q_r}}{16}} ) )
( {{\it u_3}}^{2}+{{ \it v_3}}^{2} ) ^{\frac{5}{2}}{\it v_1}) ].
\end{split}
\end{equation}
\section*{References}
\bibliography{cglbib}
\bibliographystyle{else-num}
\end{document}